\documentclass[12pt,preprint]{aastex}

\usepackage{natbib}
\usepackage{longtable}
\usepackage{rotate}
\usepackage{lscape}
\def\gtorder{\mathrel{\raise.3ex\hbox{$>$}\mkern-14mu
             \lower0.6ex\hbox{$\sim$}}}
\def\ltorder{\mathrel{\raise.3ex\hbox{$<$}\mkern-14mu
             \lower0.6ex\hbox{$\sim$}}}

\newcommand{\newQSOs}{713}	%FIXED
\newcommand{\newQSOsL}{527}	%FIXED	
\newcommand{\newQSOsS}{186}	%FIXED

\newcommand{\knoQSOs}{45}	%FIXED
\newcommand{\knoQSOsL}{38}	%FIXED	
\newcommand{\knoQSOsS}{7}	%FIXED

\newcommand{\allQSOs}{758}	%FIXED
\newcommand{\allQSOsL}{565}	%FIXED
\newcommand{\allQSOsS}{193}	%FIXED

\newcommand{\contLMC}{1017}	%FIXED
\newcommand{\featLMC}{667}	%FIXED

\newcommand{\contSMC}{344}	%FIXED
\newcommand{\featSMC}{229}	%FIXED

\newcommand{\QaA}{372}	% ALL
\newcommand{\QbA}{299}
\newcommand{\QcA}{87}

\newcommand{\QaL}{282}	% LMC
\newcommand{\QbL}{217}
\newcommand{\QcL}{66}

\newcommand{\QaS}{90}	% SMC
\newcommand{\QbS}{82}
\newcommand{\QcS}{21}

\shorttitle{MQS III. Spectroscopy of \allQSOs\ AGNs Behind the Magellanic Clouds}
\shortauthors{Koz{\l}owski et al.}

\begin{document}

\title{The Magellanic Quasars Survey. III. Spectroscopic Confirmation of \allQSOs\ AGNs Behind the Magellanic Clouds}

\author{Szymon~Koz{\l}owski\altaffilmark{1,\star},
Christopher~A.~Onken\altaffilmark{2},
Christopher~S.~Kochanek\altaffilmark{3,4},
Andrzej~Udalski\altaffilmark{1,\star}, \\
and \\
M.~K.~Szyma{\'n}ski\altaffilmark{1},
M.~Kubiak\altaffilmark{1},
G.~Pietrzy{\'n}ski\altaffilmark{1,5},
I.~Soszy{\'n}ski\altaffilmark{1},
{\L}.~Wyrzykowski\altaffilmark{1,6},
K.~Ulaczyk\altaffilmark{1}, 
R.~Poleski\altaffilmark{1,3}, 
P.~Pietrukowicz\altaffilmark{1},
and J.~Skowron\altaffilmark{1} (The OGLE Collaboration),\\
and \\
M.~Meixner\altaffilmark{7}, 
A.~Z.~Bonanos\altaffilmark{8}
}

\altaffiltext{1}{Warsaw University Observatory, Al. Ujazdowskie 4, 00-478 Warszawa, Poland; simkoz@astrouw.edu.pl}
\altaffiltext{2}{Research School of Astronomy and Astrophysics, The Australian
National University, Canberra 2611, Australia; onken@mso.anu.edu.au}
\altaffiltext{3}{Department of Astronomy, The Ohio State University, 140 West 18th Avenue, Columbus, OH
43210, USA; ckochanek@astronomy.ohio-state.edu}
\altaffiltext{4}{The Center for Cosmology and Astroparticle Physics, The Ohio State University,
191 West Woodruff Avenue, Columbus, OH 43210, USA}
\altaffiltext{5}{Universidad de Concepci{\'o}n, Departamento de Astronomia, Casilla 160-C, Concepci{\'o}n, Chile}
\altaffiltext{6}{Institute of Astronomy, University of Cambridge, Madingley Road, Cambridge CB3 0HA, UK}
\altaffiltext{7}{Space Telescope Science Institute, 3700 San Martin Drive, Baltimore, MD 21218, USA}
\altaffiltext{8}{National Observatory of Athens, Institute of Astronomy, Astrophysics, Space Applications \& Remote Sensing,
I. Metaxa \& Vas. Pavlou St., Palaia Penteli, 15236 Athens, Greece}
\altaffiltext{$\star$}{The OGLE Collaboration}

%%%%%%%%%%%%%%%%%%%%%  ABSTRACT  %%%%%%%%%%%%%%%%%%%%

\begin{abstract}
The Magellanic Quasars Survey (MQS) has now increased the number of quasars known
behind the Magellanic Clouds by almost an order of magnitude.  All survey fields in the
Large Magellanic Cloud (LMC) and 70\% of those in the Small Magellanic Cloud (SMC)
have been observed.  The targets were selected from the third phase of the Optical 
Gravitational Lensing Experiment
(OGLE-III) based on their optical variability, mid-IR and/or X-ray properties.
We spectroscopically confirmed \allQSOs\ (\allQSOsL\ LMC and \allQSOsS\ SMC) quasars behind the Clouds, of
which 94\% (\newQSOsL\ LMC and \newQSOsS\ SMC) are newly identified.
The MQS quasars have long-term (12 years and growing for OGLE), high-cadence light curves,
enabling unprecedented variability studies of quasars.  The MQS quasars also
provide a dense reference grid for measuring both the internal and bulk
proper motions of the Clouds, and 50 quasars are bright enough ($I \lesssim 18$~mag)
for absorption studies of the interstellar/galactic (ISM/IGM) medium of
the Clouds.   
\end{abstract}

\keywords{galaxies: active -- Magellanic Clouds -- quasars: general}

%%%%%%%%%%%%%%%%%%%%%  INTRODUCTION  %%%%%%%%%%%%%%%%%%%%

\section{Introduction}

The Magellanic Clouds (MCs) are the nearest well-resolved dwarf galaxies, and for decades they have 
been an ideal playground for testing stellar and galaxy evolution theories 
(e.g., \citealt{1993A&A...267..410G,2006AJ....132.2268M} and \citealt{2002AJ....124.2639V,2013ApJ...764..161K,2013arXiv1305.4641V}), 
establishing the stellar initial mass function (e.g., \citealt{1984ApJ...284..565H}), 
metallicity (e.g., \citealt{1995ApJ...438..188M}),
studying dust properties (e.g., \citealt{2001ApJ...548..296W}) 
or to test the (ultimately falsified) hypothesis of dark matter being comprised of non-luminous 
compact objects
(e.g., \citealt{2000ApJ...542..281A,2007A&A...469..387T,2011MNRAS.416.2949W}). 
Such intensively monitored areas are also ideal for finding and studying variable objects. 
For example, the third phase of OGLE
produced the OGLE-III Catalog of Variable Stars (OGLE-III CVS), 
the largest uniform catalog of variable stars in the MCs with
over 175,000 objects 
(e.g., \citealt{2008AcA....58..163S,2009AcA....59....1S,2009AcA....59..239S,2010AcA....60...17S}).
The proximity of the MCs make them well-suited to test and calibrate cosmological distance 
indicators (see \citealt{2004NewAR..48..659A} for a review), 
such as eclipsing binaries (e.g., \citealt{2011ApJ...729L...9B,2013Natur.495...76P}), 
the tip of the red giant branch (\citealt{2000A&A...359..601C, 2000AcA....50..279U}), 
Cepheids (\citealt{1999PASP..111..775F,1998ApJ...496...17G,2008AcA....58..163S}),  
RR Lyrae (\citealt{2000AcA....50..279U}), 
cluster main-sequence fitting (\citealt{1984ApJ...285L..53S}), or 
red clump stars (\citealt{2000ApJ...531L..25U, 2002ApJ...573L..51A}). 

There are, however, more uses for such a huge database of photometric records collected in this region 
of the sky.  With little Galactic or MC extinction, it is in principle straightforward to find supernovae 
(e.g., \citealt{2013AcA....63....1K}) and active galactic nuclei (AGNs\footnote{AGNs, quasars, and QSOs 
will be used interchangeably throughout this paper.}) behind the MCs.  The challenge, of course,
is that while there are $\sim$25~quasars/deg$^2$ with $\hbox{I} < 20$~mag, there are over $10^6$
stars/deg$^2$ in the MCs.  With so many stars of different types, optical color selection methods
have too high a false positive rate.  Wide area X-ray surveys suffer both from contamination by
accreting sources in the MCs and low ($\sim$~arcmin) resolution that makes it difficult to correctly
identify the optical counterpart.   Early searches based on variability lacked robust, quantitative
means of distinguishing the aperiodic variability of quasars and stars.  Despite these difficulties,
some $\sim$80 quasars had been discovered behind the MCs as of 2009
(\citealt{1994PASP..106..843S,2002ApJ...569L..15D,2003AJ....125.1330D,2003AJ....126..734D,2003AJ....125....1G,2005A&A...442..495D}), 
and they were a crucial part of the projects to accurately measure the proper
motions of the MCs (\citealt{2006ApJ...638..772K,2006ApJ...652.1213K,2008AJ....135.1024P,2013ApJ...764..161K}).  
Now, however, these projects are limited by the lack of a denser reference
grid that can be used to better measure and separate the internal and bulk motions
of the MCs (e.g., \citealt{2013ApJ...764..161K}).  However, expanding the quasar sample 
by an order of magnitude was probably infeasible using the approaches of these earlier searches.  

The first major improvements became possible with the advent of the {\it Spitzer Space Telescope} 
(\citealt{2004ApJS..154....1W}).  It allowed mid-IR surveys of the MCs such as 
the {\it Surveying the Agents of Galaxy Evolution} (SAGE, \citealt{2006AJ....132.2268M}), 
{\it Surveying the Agents of Galaxy Evolution--SMC} (SAGE--SMC, \citealt{2011AJ....142..102G}),
and {\it Spitzer Survey of the Small Magellanic Cloud} 
(S$^3$MC, \citealt{2007ApJ...655..212B}) projects.  At the same time, it was realized in 
extragalactic surveys that mid-IR colors were a powerful means of distinguishing stars,
galaxies and AGNs -- in particular, almost all red mid-IR sources are quasars because 
they have a flatter spectral energy distribution than the Rayleigh-Jeans law that
(roughly) characterizes stars and low redshift galaxies
(\citealt{2004ApJS..154..166L,2005ApJ...631..163S}).  In \cite{2009ApJ...701..508K}, we
showed that this was also true in the dense stellar fields of the MCs, particularly
with the addition of limits on the OGLE-III (\citealt{2008AcA....58...89U, 
2008AcA....58..329U, 2008AcA....58...69U}) $I$-band-to-mid-IR colors, albeit with some additional
contamination from the higher abundances of dusty stars and young stellar objects (YSO). 

At the same time, \cite{2009ApJ...698..895K} proposed that quasar light curves were well
modeled by a stochastic process, the damped random walk (DRW), which is characterized by
an exponential covariance matrix defined by an asymptotic variance $\sigma$ and a time
scale $\tau$.  In \cite{2010ApJ...708..927K}, we showed that the mid-IR quasar candidates
from \cite{2009ApJ...701..508K} largely lay in a different region of the  $\sigma$/$\tau$
parameter space from variable stars, thus providing a robust, quantitative means of 
variability-selecting quasars.  \cite{2010ApJ...708..927K} also (re)introduced a more
statistically powerful method of estimating the DRW parameters than used by
\cite{2009ApJ...698..895K}, based on the methods previously discussed
by \cite{1992ApJ...385..404P} and \cite{1992ApJ...398..169R,1994comp.gas..5004R}.  This 
was then confirmed by \cite{2010ApJ...721.1014M} using the SDSS variability 
data on $\sim$9000 spectroscopically confirmed quasars in SDSS Stripe 82 AGNs
with extensions by \cite{2011ApJ...728...26M} and \cite{2011AJ....141...93B}.
Other recent studies on variability-selecting
quasars can be found in \cite{2002AcA....52..241E},
\cite{2010ApJ...714.1194S}, \cite{2011A&A...530A.122P}, \cite{2012ApJ...747..107K}, 
and \cite{2012MNRAS.427.1284P}.

\cite{2009ApJ...698..895K} and \cite{2010ApJ...721.1014M} further showed that the DRW
parameters are correlated with wavelength, luminosity and black hole mass, and 
\cite{2012ApJ...753..106M} show that it can fully explain the variability statistics 
of ensembles of quasars. \cite{2013ApJ...765..106Z} and \cite{2013arXiv1304.2863A}
show that on time scales of days to years the DRW model is a better model stochastic
process for quasar light curves than many simple variants, although \cite{2011ApJ...743L..12M}
found for four Kepler-monitored AGNs that the power spectrum may steepen on very short
time scales.  The DRW model then provides a very well-defined means of carrying out
the interpolations needed when cross-correlating light curves, as shown in the
reanalysis of quasar reverberation mapping light curves by \cite{2011ApJ...735...80Z},
and it could play a similar role in measuring the time delays of lensed quasars
(e.g., \citealt{1992ApJ...385..404P}, \citealt{2013arXiv1304.0309H}, who use a stochastic
model but not the DRW model).

A fundamental problem with this renaissance in quasar variability studies is that
the SDSS Stripe 82 light curves are not, in fact, very good, comprising only 
60 epochs for each quasar with large temporal gaps. 
In fact, the quasars with the best, densely-sampled, long-term light curves, are
the quasars behind the MCs, because they have been almost continuously monitored
by microlensing projects for over a decade.  The typical quasar has  
$\sim$500 $I$-band and $\sim$50 $V$-band
epochs from OGLE-III (years 2001-2009) and $\sim$500 $I$-band and $\sim$100 $V$-band
epochs from OGLE-IV (years 2010-2013) and the light curves continue to be extended.
The superiority of these light curves is likely to remain the case
until 10 years after the advent of LSST (\citealt{2008arXiv0805.2366I}), 
although Pan-STARRS (\citealt{2002SPIE.4836..154K}) will provide 
larger numbers of more sparsely sampled, multi-color light curves.  As noted
earlier, denser networks of quasars behind the MCs are also needed for improved
proper motion measurements.      
  
We started the {\it Magellanic Quasars Survey} (MQS) in 2009 
to greatly expand the number of AGNs behind the MCs using the 3.9 meter 
Anglo-Australian Telescope (AAT) and the AAOmega spectrograph.
AAOmega allows multi-object spectroscopy of $400$ targets within a 3 deg$^2$ 
field of view (e.g., \citealt{2006SPIE.6269E..14S}).  Although the runs 
were plagued by bad weather, we reported the discovery of $29$ new AGNs behind 
the SMC (doubling their number) in \citeauthor{2011ApJS..194...22K} 
(\citeyear{2011ApJS..194...22K}, hereafter Paper~I) and the discovery of 
$144$ new AGNs behind the LMC (quadrupling their number) in \citeauthor{2012ApJ...746...27K} 
(\citeyear{2012ApJ...746...27K}, hereafter Paper~II).  Here, we bring the present phase
of the MQS (spectroscopic confirmations) to a conclusion, where we 
have completed all 12 of the planned LMC fields and
three out of five of the planned SMC fields (plus the pilot study from Paper I), 
confirming a total of \allQSOs\ AGNs, which represents an increase in the
number of known quasars behind the MCs by almost an order of magnitude. 
In Section~\ref{sec:AGNsel}, we describe the AGN selection procedures, and
in Section~\ref{sec:data} we describe the data and their analysis.
New AGNs are presented in Section~\ref{sec:newquasars}, and the contaminating
objects and objects with featureless spectra are described in Section~\ref{sec:contamination}.
In Section~\ref{sec:discussion}, we discuss the relative detection efficiencies
of our AGN selection methods.
The paper is summarized in Section~\ref{sec:summary} and we outline future areas
of exploration.

%%%%%%%%%%%%%%%%%%%%%  AGN SELECTION  %%%%%%%%%%%%%%%%%%%%

\section{AGN Selection Procedures}
\label{sec:AGNsel}

Our main driver for finding AGNs behind the MCs was to study their variability.
We therefore limited our search to the well-monitored OGLE-III fields (\citealt{2008AcA....58...89U, 
2008AcA....58..329U, 2008AcA....58...69U}). We cross-matched the OGLE data
with the SAGE, SAGE-SMC and S$^3$MC mid-IR data (\citealt{2006AJ....132.2268M,2011AJ....142..102G,2007ApJ...655..212B}) and 
the ROSAT X-ray catalogs (\citealt{1999A&AS..139..277H,2000A&AS..142...41H}).  The exact selection 
procedures are outlined in Paper~II. For completeness, we briefly sketch these procedures here.

{\it Method~1.} In the first method, we use the mid-IR/optical color-selected AGN candidates 
from \cite{2009ApJ...701..508K}.  In the mid-IR color-color space we defined a wedge
following \cite{2005ApJ...631..163S} that we further subdivided into region A, which should
be free of low temperature (dusty) black bodies, and region B which might contain them.
Then in the mid-IR color-magnitude diagram (CMD), we defined a region likely to be heavily
contaminated with young stellar objects (YSO) and one which should mostly contain quasars
(QSO).  Finally, we defined objects with mid-IR-to-optical colors similar to AGNs as class
``a'' and those with other colors as class ``b''.  Hence, each object has a classification such
as QSO-Aa (most pure), QSO-Ab, $\cdots$, YSO-Ba,  or YSO-Bb (most contaminated). Where 
there was no optical match we assigned a classification of (Q/Y)SO-(A/B)0.

{\it Method~2.} Our second criterion was variability, and the original intent was simply to
use our criteria from \cite{2010ApJ...708..927K}.  At the time, however, there was an
unresolved problem in the long term OGLE-III light curves involving inter-seasonal jumps in
the photometry that then triggered large numbers of false positives when we simply fit the
DRW model to all available light curves.  Lacking time to resolve this problem for the full
$\sim 42$~deg$^2$ survey area, we largely
adopted a variant of the \cite{2010ApJ...714.1194S} structure function selection method
to screen candidates because it was less sensitive to localized jumps.  The resulting
cuts were as follows (as in Paper~II):  

\noindent {\it Cut 1.} The average light curve magnitude is $16.0<I<19.5$~mag for the LMC 
and $16.5<I<19.5$~mag for the SMC. The faint limit ensures that the data has a high enough 
signal-to-noise ratio to provide a good light curve and the bright limit eliminates 
variable stars at fluxes where we have no significant expectation of finding a 
quasar given the survey area (see \citealt{2010ApJ...708..927K});

\noindent {\it Cut 2.} The light curve must be fit by some DRW model better than it
   is fit by white noise ($\ln L_{\rm best}>\ln L_{\rm noise}+2$, \citealt{2010ApJ...708..927K}).
   This essentially selects sources which are variable due to something other than noise;

\noindent {\it Cut 3.} We applied a very loose constraint on the DRW model time scale
  ($1 < \log(\tau/{\rm days}) < 5$) and no constraint on $\hat{\sigma}$. The restrictions
  on $\tau$ removed 62\% of the sources, mostly at short time scales. 

\noindent {\it Cut 4.} The slope of the light curve's structure function was broadly consistent
   with that of a quasar ($0.1<\gamma<0.9$, see \citealt{2010ApJ...714.1194S});

\noindent {\it Cut 5.} Finally, we also limited the associated $I$-band amplitude of the 
   structure function to $A<0.4$~mag to remove high amplitude variable stars;

\noindent In an ideal world we would have used a ``cleaner'' procedure so that our final discussion
of the variability selection results would be simpler.  On the other hand, we had no shortage of
fibers, so there was no harm in using a rather broad definition for variability-selected candidates. 
In Section~\ref{sec:discussion}, we also comment on results when applying the original 
\cite{2010ApJ...708..927K} variability criteria.

{\it Method~3.} Finally, we included variable objects with positions consistent with the location of
any ROSAT X-ray source.  Where there were multiple variable sources, the one closest to the
X-ray position was included. 

The AAT/AAOmega {\sc configure} software (\citealt{2002MNRAS.333..279L,2006MNRAS.371.1537M}) allows 
assigning priorities to targeted objects, where priority 1 is the lowest and 9 is the highest. 
We prioritized our candidates in the following way. Priority 9 was assigned to objects selected by 
all three methods, priority 8 was assigned to objects meeting any two selection criteria, and 
priority 7 was assigned to objects selected by a single method.  With the availability of fibers, 
we also included 931 stars (286 observed) that could potentially have been stripped from the 47 Tuc 
globular cluster (\citealt{2012MNRAS.423.2845L}). These stars should be easy to kinematically
separate from SMC stars even with our relatively low spectral resolution.  
These stars were assigned priority 6, and the results will be presented elsewhere.

It is important to realize that we are not trying to produce
a very high purity candidate sample because AAOmega has a significantly higher density of 
fibers (400/3~deg$^2\approx130$/deg$^2$) than there are $I<20$~mag quasars ($\sim 25$/deg$^2$) for which 
we are likely
to measure redshifts given the exposure times and the backgrounds created by the high stellar 
density and large aperture fibers.  Every fiber is ultimately assigned to something that
might be a candidate (modulo the 47 Tuc stars), although many will be in low purity 
sub-samples (e.g., YSO-Bb) or so faint that we will only obtain
a redshift if the source has sufficiently strong lines.  Contamination  is particularly severe
for the very brightest (and rarest) quasars which both have best light curves and are the 
most useful for any absorption line studies.  By definition,   only 1/4-1/3
of our targets can be quasars bright enough to measure a redshift, which means we will
also produce a large spectroscopic catalog of
dusty or otherwise peculiar stars as a consequence. The stellar content of the MQS
will be considered elsewhere.

%%%%%%%%%%%%%%%%%%%%%  DATA  %%%%%%%%%%%%%%%%%%%%

\section{Data}
\label{sec:data}

The three selection criteria lead to samples of 2434 and 1447 
candidates behind the OGLE-III regions of the  
LMC and SMC.  For completeness, we include all MQS sources in our summary tables,
but for discussions of efficiencies we exclude the sources from Paper I (where
we used somewhat different selection methods) that did not pass our selection 
criteria from Section~\ref{sec:AGNsel}.  The basic information on the
fields is provided in Table~\ref{tab:MQSlog} and their locations on the 
sky are shown in Figure~\ref{fig:MQS_fields}.  The target integration time
was 1.5 hours ($3 \times 30$~minutes), and this was obtained for 13 of the
15 completed fields (Table~\ref{tab:MQSlog}).  Three of the fields from
Paper II had shorter than desired exposure times, but we repeated one of
these fields (LMC4a as field LMC4b) during the final observing run.

We used the 580V (blue channel) and 385R (red channel) gratings to provide a
resolution of $R \approx 1300$ and a spectral range of 3700\AA--8800\AA, with
the spectra spliced near 5700\AA.  This broad coverage, low resolution mode is
well-suited for AGN identification since we are interested in relatively broad,
but sparse lines.  The data were reduced and calibrated with the standard AAOmega {\sc 2dfdr} routines 
(\citealt{1996ASPC..101..195T}).  We then inspected all the spectra using our own 
dedicated code for finding AGNs. We searched for the common redshifted AGN spectral lines 
(see e.g., \citealt{2001AJ....122..549V}) such as hydrogen Ly$\alpha$ at 1216\AA, H$\delta$ at 4101\AA, 
H$\gamma$ at 4340\AA, H$\beta$ at 4861\AA, H$\alpha$ at 6563\AA, magnesium
MgII at 2800\AA, carbon CIV at 1549\AA~and CIII] at 1909\AA, and also 
the narrow lines of oxygen [O II] at 3727\AA, [O III] at 4959\AA~and 5007\AA.
The AGN identification was viewed as confirmed if at least two AGN lines were identified,
with the exception of the redshift range from $0.7<z<1.2$ where MgII is frequently the 
only observable line. We paid special attention to $z\approx1$ AGNs, for which incorrect splicing
of blue and red spectra at 5700\AA\ can mimic MgII line.

%%%%%%%%%%%%%%%%%%%%%%%%%%%%%%%%%%%%%%%%%%%%%%%%%%%%%%%%%%%%%%%%%%%%%%%%%%%%%%%%%%%

\section{New Quasars}
\label{sec:newquasars}

We identified a total of \allQSOs\ quasars, \allQSOsL\ in the LMC and \allQSOsS\ in the SMC, from our 
targeted sample of 2248 LMC  and 766 SMC sources, including those reported earlier in Papers~I and II.
We chose targets independent of any prior identifications. Of the 66 known AGNs in our observed fields, we
selected 48 as candidates, observed 46 and re-confirmed \knoQSOs\ of them 
(we discuss the exception below), leaving a total of \newQSOsL\ and \newQSOsS\ new MQS quasars behind the LMC 
and SMC, respectively.  Of the 18 known AGN that were not selected as candidates, five were in the
pilot field of Paper I that was governed by a different set of selection criteria 
(although not observed, three of them were included in the current candidate list, the 
remaining two had incomplete mid-IR colors, no X-ray emission and insignificant optical variability).
Of the 13 known AGN in the observed standard fields, six lacked (complete) Spitzer photometry, 
six were not detected by Spitzer and one lay outside the mid-IR selection region.
Most (nine) were not significantly variable, although we did lose one known AGN for being
``too variable'' and having a structure function amplitude $A>0.4$~mag. 
Three of them were variable but fainter than $I>19.5$~mag. All but three had no X-ray counterparts, 
and these three remaining sources had ROSAT detection probabilities below the threshold we used for our 
target selection.  

\begin{figure*}
\centering
\includegraphics[width=11cm]{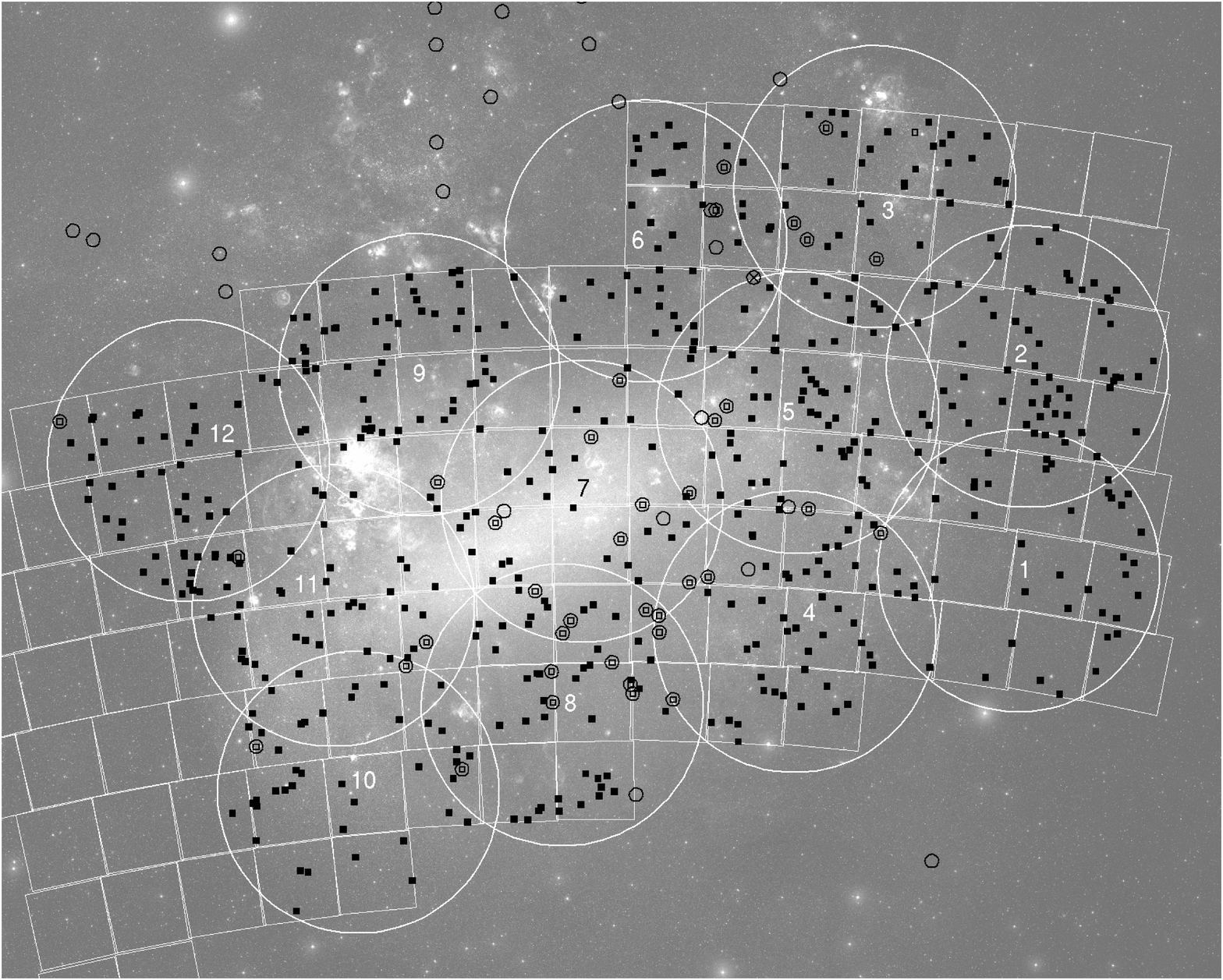}\\
\vspace{0.1cm}
\includegraphics[width=11cm]{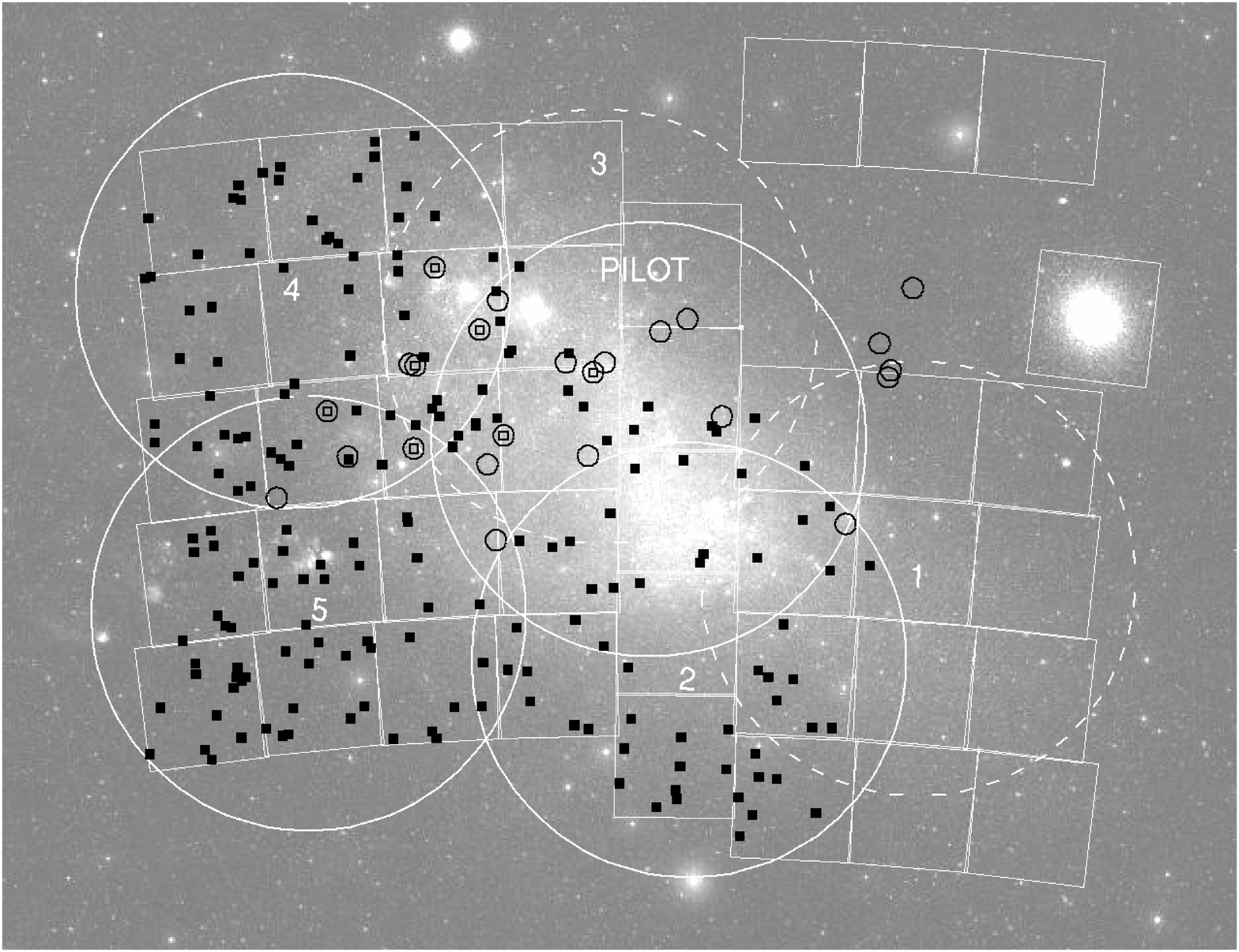}\\
\caption{The twelve MQS LMC (top) and six SMC (bottom) fields identified by
their field numbers.  In addition to the three official SMC fields that were observed 
(solid lines; fields 2, 4, and 5), we show our pilot field and the two unobserved SMC
fields as dashed circles.  Black squares mark our new MQS confirmed AGNs, open circles previously known quasars, which then contain
a central open square if they were reobserved and confirmed.  The $\otimes$ symbol 
(near the overlap of the LMC~3, 5, and 6 fields) marks the one
exception, the AGN from \cite{2005A&A...442..495D} that we were unable to confirm.  The white
squares outline the  OGLE-III fields.  The top (bottom) image covers approximately 
$9^\circ\times7^\circ$ ($5^\circ\times4^\circ$). 
North is up, east is to the left.}
\label{fig:MQS_fields}
\end{figure*}

\begin{figure*}
\centering
\includegraphics[width=12.5cm]{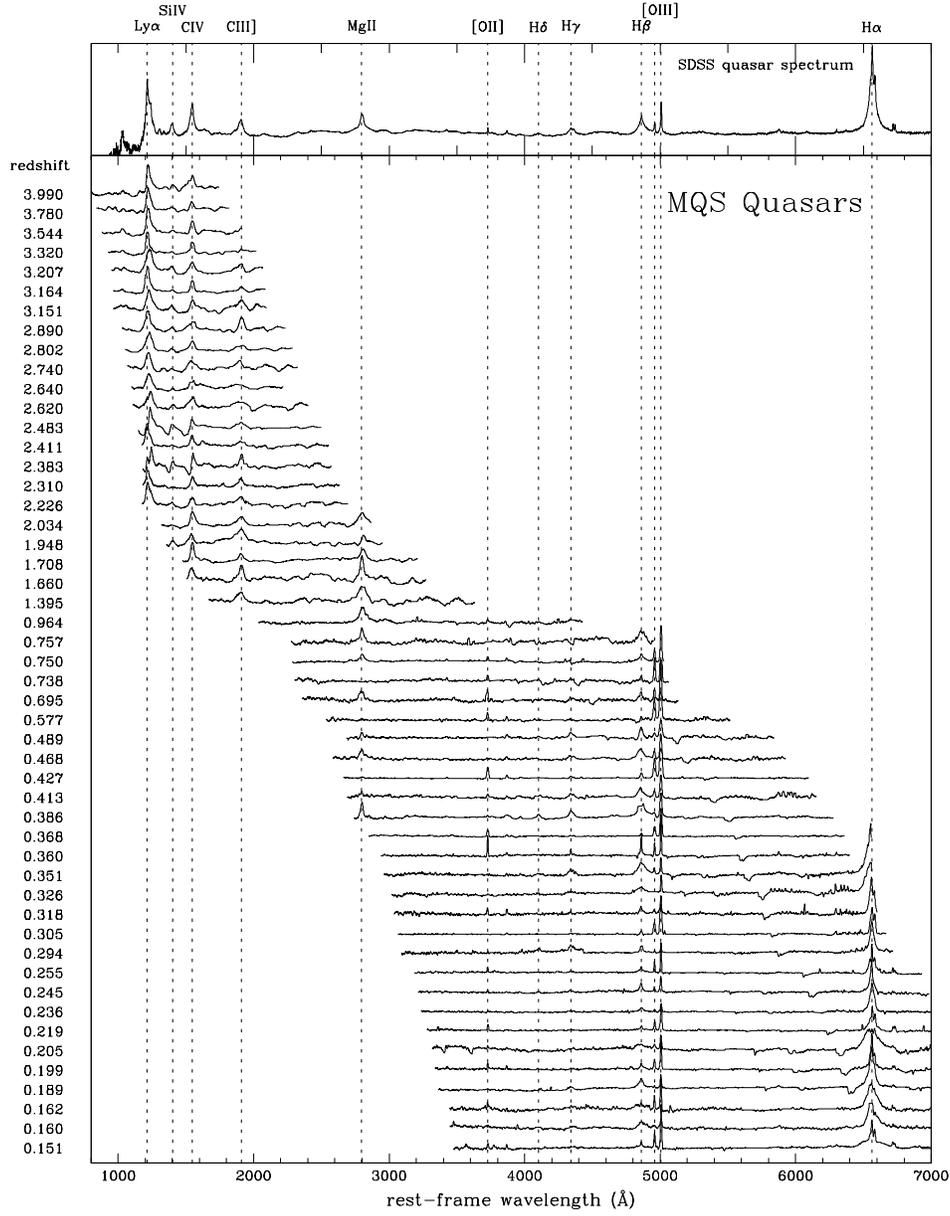}
\caption{{\it Main panel:} Rest frame spectra of 50 out of the \newQSOs\ new LMC/SMC AGNs reported in this paper. 
The spectra have been flattened, smoothed and scaled. 
The majority of the $z\approx0$ LMC/SMC emission lines as well as some of the atmospheric 
absorption features have been masked in order to emphasize the quasar emission lines. 
Each spectrum is labeled by redshift (on the left) and we also mark the common quasar 
lines (vertical dashed lines with labels).
{\it Top panel:} For comparison we show the composite quasar spectrum 
(detrended, flattened, and scaled) based on 2200 spectra from SDSS (\citealt{2001AJ....122..549V}).}
\label{fig:spectra}
\end{figure*}

The location of new (and previously known) AGNs on the sky 
is shown in Figure~\ref{fig:MQS_fields}, while their basic properties and coordinates are reported 
in Tables~\ref{tab:results_LMC} (LMC) and \ref{tab:results_SMC} (SMC).  Selected spectra of 50 new 
MQS AGNs are shown in Figure~\ref{fig:spectra}.  All identified AGNs had their spectra classified 
into quality classes: Q1 for obvious AGN spectra, Q2 for relatively obvious AGN spectra with 
problems/contamination, and Q3 for just above a borderline, usually low S/N or highly contaminated 
AGN spectra. There are \QaA\ (\QaL, \QaS) Q1 
AGNs behind the MCs (LMC, SMC), \QbA\ (\QbL, \QbS) Q2, and \QcA\ (\QcL, \QcS) Q3.
For sources brighter than $I<19.5$ mag, 58\% of them are Q1 AGNs, while for sources fainter than 
this limit only 35\% are Q1, simply reflecting decreasing spectra quality with decreasing S/N.

One AGN from Paper II, AGN MQS~J051509.61$-$701711.7, turned out to be a false positive, where we 
mis-identified the rest-frame [OI] lines at 6300\AA\ and 6364\AA\ as the [OIII] 4959\AA~and 5007\AA\ 
lines at a redshift of $z\approx0.27$, and therefore has been deleted from the final sample.  We 
inspected all other AGNs at similar redshifts and found no other mis-identifications. We were 
unable to confirm the AGN J050550.35$-$675017.5 from \cite{2005A&A...442..495D}.  They
selected this as an X-ray source from higher resolution (than ROSAT) XMM-Newton data, while
we selected it as a QSO-Aa mid-IR candidate.  In the OGLE-III images it is associated with
a $\sim 13 \times 3$~arcsec, mildly edge-on galaxy that may have a bright nucleus.  As such,
the source is almost certainly an AGN.  However, \cite{2005A&A...442..495D} assign
a redshift of a $z=0.07$ quasar based on a single noisy line interpreted as H$\alpha$, 
and we are unable to confirm this redshift or identify an alternative.  We count this source
as unconfirmed in our statistical discussions.  
This source is marked with a ``$\otimes$'' symbol in 
Figure~\ref{fig:MQS_fields}. There are no high resolution images available from the {\it Hubble 
Space Telescope} (HST) archives, so it was not used in any of the HST proper motion studies.

\begin{figure*}
\centering
\includegraphics[width=12.5cm]{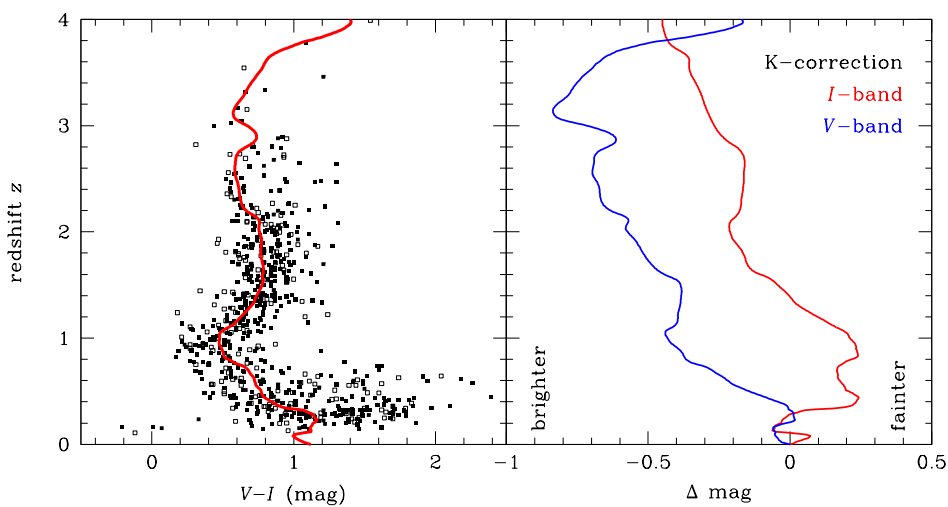}
\caption{AGN colors (left) and $K$-corrections (right) as a function of redshift $z$. 
{\it Left panel:} Filled (open) squares represent the LMC (SMC) quasars confirmed by MQS. 
The red line is expected $V-I$ color derived from the average SDSS quasar spectrum
from \cite{2001AJ....122..549V} as it is redshifted through the OGLE filters. Significant outliers 
from this line are AGNs blended with stellar light from the MCs. Lower luminosity AGN
at lower redshifts are frequently redder because of increased contamination from their host 
galaxies.  {\it Right panel:} AGN $K$-corrections for the $V$- (blue) and $I$-band (red) 
OGLE filters.} 
\label{fig:QSO_color}
\end{figure*}

\begin{figure}
\centering
\includegraphics[width=12.5cm]{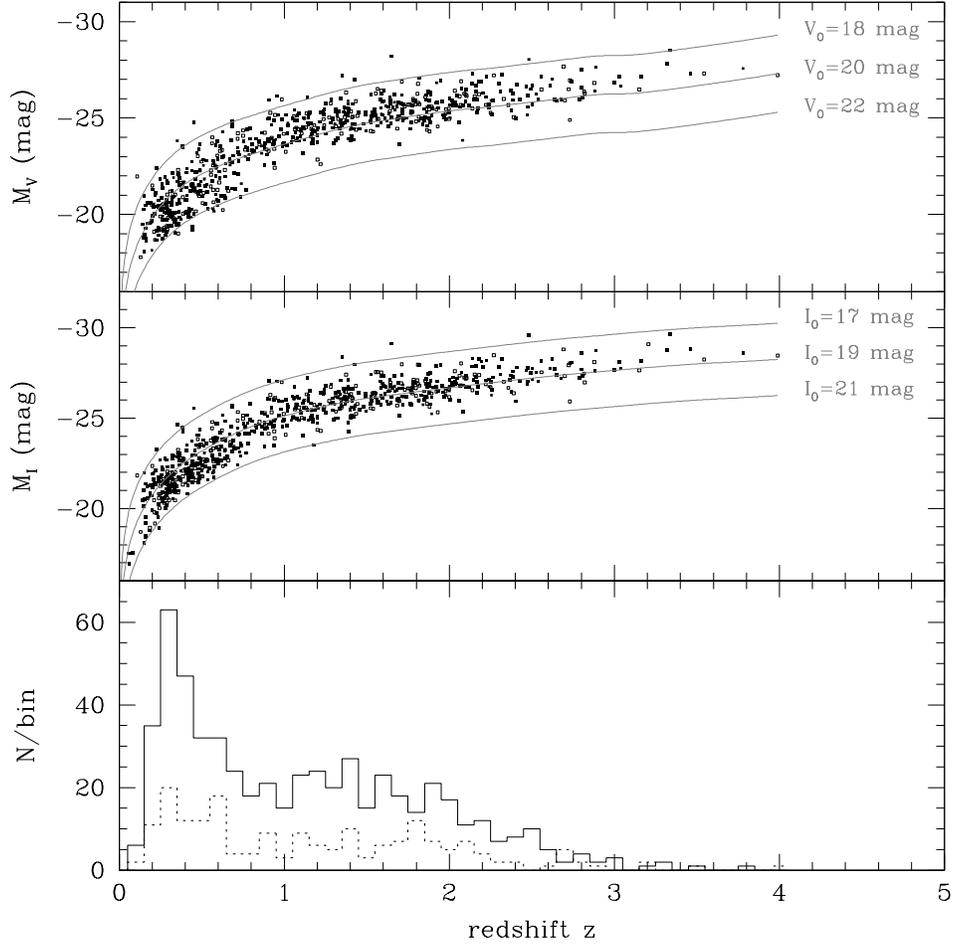}
\caption{Absolute magnitudes of MQS AGNs in $V$-band (top panel) and $I$-band (middle). 
Solid (open) squares are for the LMC (SMC) AGNs.
We also show lines of constant observed (extinction-corrected) magnitude.
The bottom panel shows redshift histograms of the confirmed quasars behind
the LMC (solid) and the SMC (dotted).}
\label{fig:hist}
\end{figure}

\begin{figure}
\centering
\includegraphics[width=12.5cm]{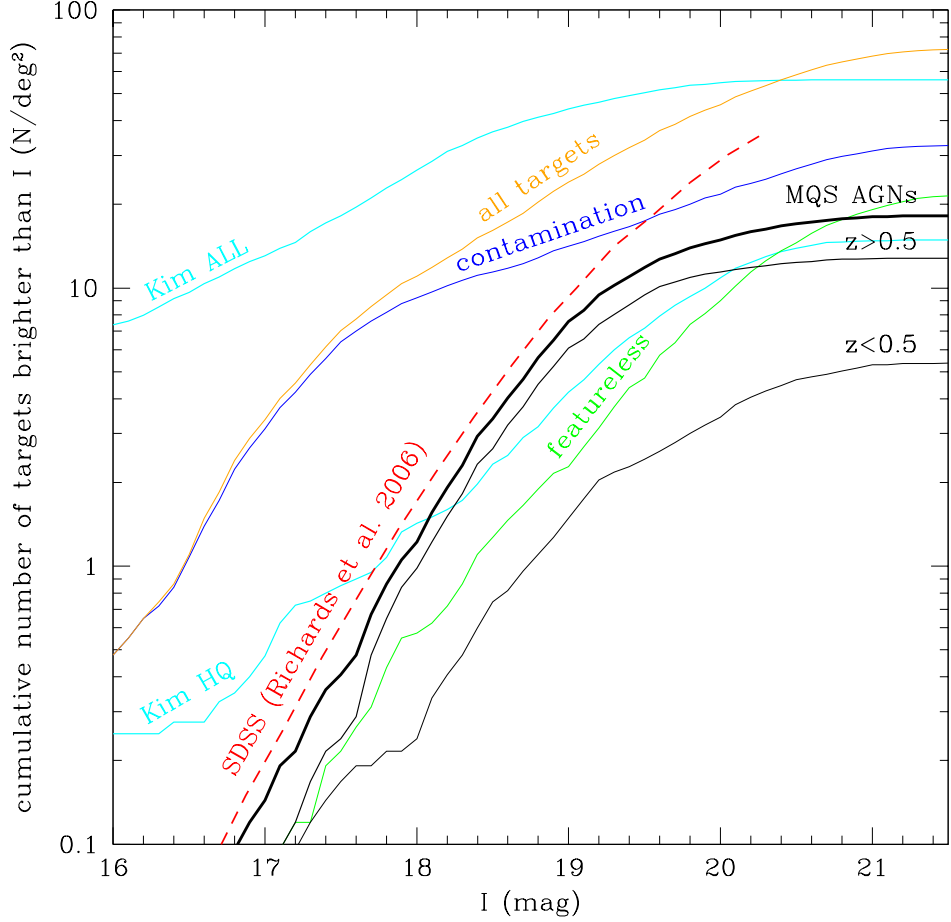}
\caption{Cumulative distribution of MQS AGNs above a given $I$-band magnitude (thick black line)
as compared to SDSS quasars (\cite{2006AJ....131.2766R}, converted to $I$-band; red).  The MQS survey
is roughly $\sim$75\% complete for $I<19$~mag.  We also show the cumulative distributions of
$z<0.5$ and $z>0.5$ MQS AGNs (narrow black lines), all targets (orange), contaminating (blue) and 
featureless sources (green). Finally, we show the distributions of the full (Kim ALL) and ``high quality'' (Kim HQ) samples from
\cite{2012ApJ...747..107K} in cyan. To put these surface densities in perspective, the density of
AAOmega fibers ($\sim 130$~fibers/deg$^2$) lies above the upper scale of the figure.}
\label{fig:hist_cumul}
\end{figure}

Figures~\ref{fig:QSO_color}-\ref{fig:hist_cumul} summarize various properties of the sample. 
Figure~\ref{fig:QSO_color} shows the distribution of observed and expected optical colors
as a function of redshift, where we compute the expected colors and $K$-corrections using
the template AGN spectrum from \cite{2001AJ....122..549V}.  We use Galactic/MC extinction corrections
 from \cite{2011AJ....141..158H}.  We combine the $K$-corrections and the extinction 
to estimate the absolute magnitude of each AGN assuming a standard $\Lambda$CDM cosmological model with 
($\Omega_\Lambda$, $\Omega_M$, $\Omega_k$) = (0.7, 0.3, 0.0) and h = $H_0$/(100 km/s/Mpc) = 0.71 to 
calculate luminosity distances.  Figure~\ref{fig:hist} shows the distribution of the AGN
in absolute $V$- (top) and $I$-band (middle) magnitudes along with a histogram of the
overall redshift distribution (bottom).  Finally, Figure~\ref{fig:hist_cumul} shows the
cumulative surface density of the sample as a function of $I$-band magnitude. If we
compare this to the SDSS $i$-band number counts from \cite{2006AJ....131.2766R}, 
corrected to the OGLE $I$-band (shifted by $-$0.3 mag), we see that the MQS sample is roughly 
$\sim$75\% complete 
for $I<19$~mag, which seems quite good given the nature of the survey fields!  Some of
the incompleteness is associated with regions of very high stellar density, as illustrated
by the lower number of quasars directly behind the central regions of the MCs.

%%%%%%%%%%%%%%%%%%%%%%%%%%%%%%%%%%%%%%%%%%%%%%%%%%%%%%%%%%%%%%%%%%%%%%%%%%%%%%%%%%%

\section{Unidentified and Contaminating Sources}
\label{sec:contamination}

The remaining LMC (SMC) sources can be divided into \contLMC\ (\contSMC) contaminating 
sources and \featLMC\ (\featSMC) objects with featureless spectra, where a contaminating
source is clearly some sort of stellar source in the LMC and a featureless spectrum
is one where the S/N is simply too poor to propose a classification.
In Paper~II, we investigated the nature of the contaminating sources and found that they
are typically planetary nebulae (PNe), YSOs, B/Be stars, etc., as might be expected
from the requirement that they show dust emission, variability or X-ray emission.  
The properties of the final larger sample will be explored elsewhere.  
Figure~\ref{fig:hist_cumul} compares the cumulative distributions of these sources
to that of the AGN.  We see that contaminating sources dominate the overall
target distribution at bright magnitudes and that featureless sources dominate
at faint magnitudes.   Essentially, filling the fibers means we can look at 
all possible bright candidates and gamble that we might identify quasars fainter
than $I\sim20$~mag despite the high effective sky backgrounds. This leads to 
a low apparent detection efficiency of $\sim$30\%, 
but is really just a consequence of using all available fibers. 

%%%%%%%%%%%%%%%%%%%%%  DISCUSSION  %%%%%%%%%%%%%%%%%%%%

\section{AGN Selection Methods}
\label{sec:discussion}

We can use the overall sample to explore the various search methods proposed
to identify quasars behind the MCs.   In some sense, this question is almost
moot, since the MQS has already identified the majority of bright quasars
behind the densest regions of the MCs, and the problem becomes simpler
in any expansion of the search region because the stellar densities are lower.
These issues would be relevant, however, to attempts to find fainter quasars,
although there is no immediately obvious scientific driver for such a search.
Table~\ref{tab:selres} summarizes the statistics for the various methods,
where readers should focus on the differences in efficiencies rather than
the absolute efficiencies since the latter are by definition low because
of our strategy of using every fiber.

For the present analysis we will discuss relative completenesses more carefully
than in Paper~II.  The extra complication is that we assigned quasars an
observational  priority based on whether they were selected based on $i=1$,
$2$, or $3$ methods, so the fraction observed $f_i$ depends on $i$.  For $i=1$ and
$2$ we can assume that the probability of being observed was independent of
which methods identified the candidate since that information was not used
in setting the priorities.  For any particular class of objects (e.g., QSO-Aa)
there were then $N_i$ candidates yielding $Q_i$ quasars, so the overall
efficiency for the class is
\begin{equation}
\label{eqn}
      E = \left[ \sum_i Q_i f_i^{-1} \right] \left[ \sum_i N_i \right]^{-1}.
\end{equation}
Note that the total number of candidates is $N_i = O_i/f_i$ where $O_i$
is the number of candidates that were observed, so if all priorities were 
observed with equal probability ($f_1=f_2=f_3$) the efficiency is simply
the number of quasars found divided by the number of objects observed. 

\begin{figure*}
\centering
\includegraphics[width=12.5cm]{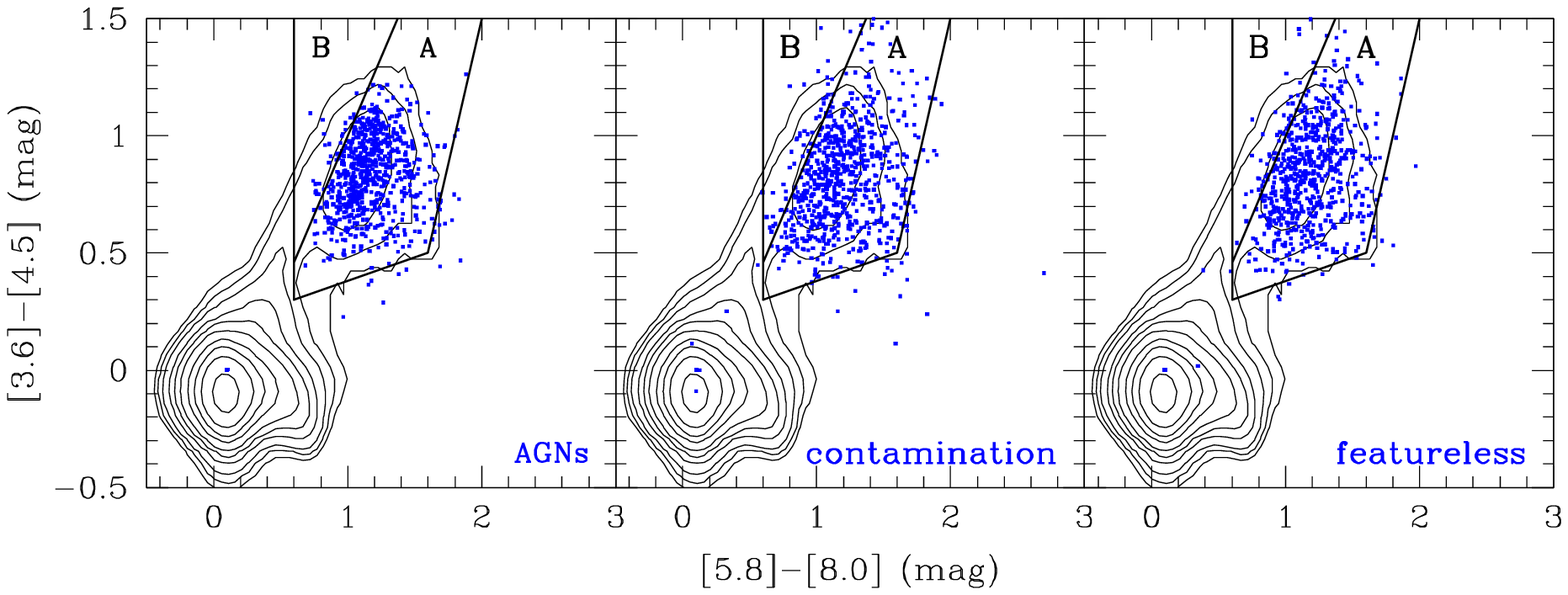}\\
\vspace{0.3cm}
\includegraphics[width=12.5cm]{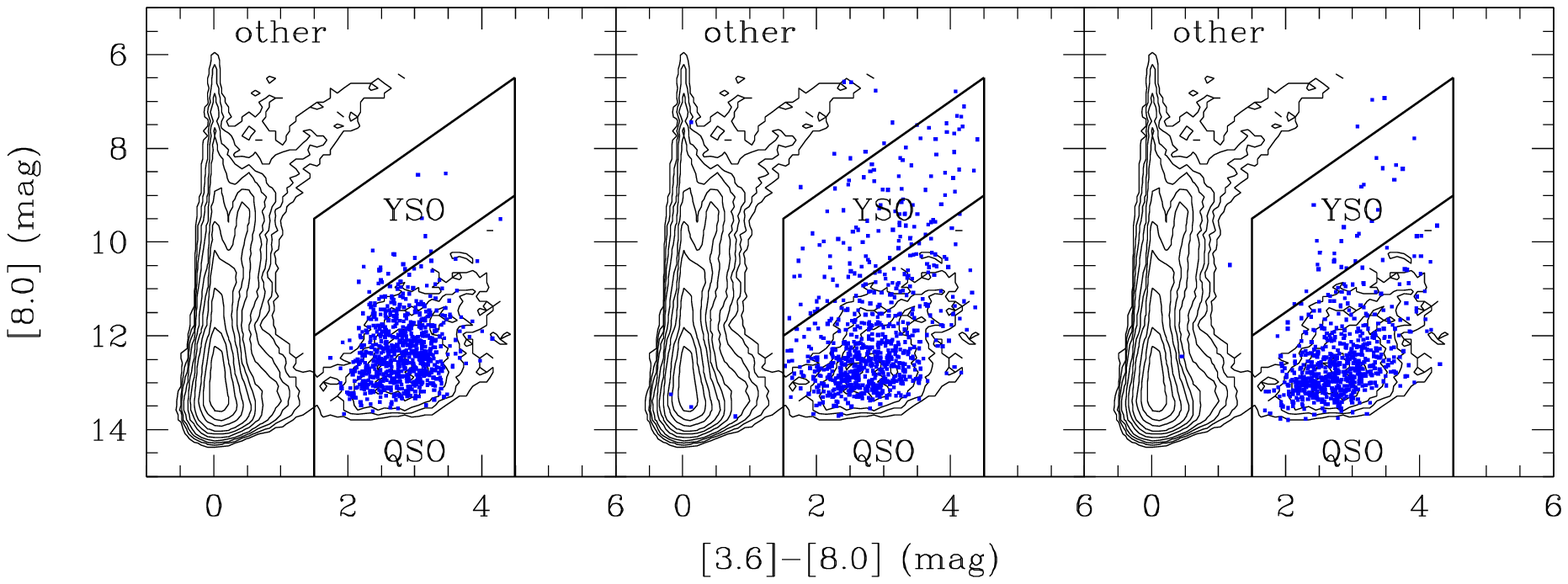}\\
\vspace{0.3cm}
\includegraphics[width=12.5cm]{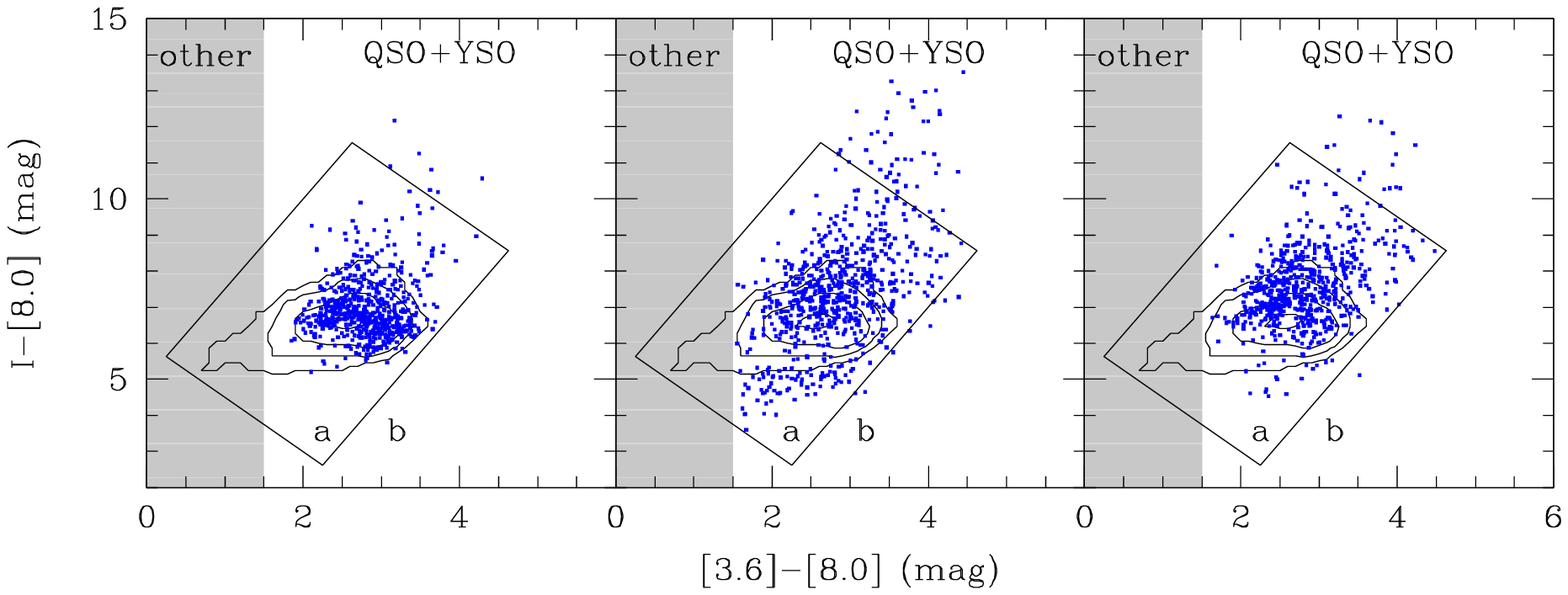}
\caption{The mid-IR AGN selection criteria from \cite{2009ApJ...701..508K}.  The points 
show the distribution of all MQS quasars (left panel), contaminating sources (middle) and 
featureless sources (right), and the labeled regions show the mid-IR/optical selection
regions.   Sources outside the selection regions did not meet the mid-IR selection
criterion but were either variability- and/or X-ray-selected.  The contours
show the distribution of all SAGE (\citealt{2006AJ....132.2268M}) sources in the LMC.
}
\label{fig:KK09cuts}
\end{figure*}

Figure~\ref{fig:KK09cuts} shows the mid-IR selection criteria we introduced
in \cite{2009ApJ...701..508K}.  Table~\ref{tab:selres} summarizes the various mid-IR
selection groups, where in our discussion we will ignore those
with few observed sources (like YSO-Ba). As expected, the highest yield is for QSO-Aa objects 
($\sim$29\%) followed by QSO-Ba ($\sim$24\%).  Much of this is driven by
our inclusion of faint sources, and if we restrict the sample to $I<19.5$~mag 
(the bright sample, hereafter) the efficiency rises to 49\% and 51\%, respectively.  
As expected, the YSO regions have lower 
yields ($\sim$20\%), and the yields become very low ($<10\%$) if a target did not have
the typical optical/mid-IR color of quasars in the AGES (\citealt{2012ApJS..200....8K}) 
survey (class ``b'' rather than ``a'').  
It appears that the distinction between sources along the
black-body color track (class B) as compared to those off that color track (class A)
has little effect and could simply be dropped.  Overall, the yield for a source
satisfying any of the mid-IR criterion was 27\% for all sources and 44\% for 
bright sources (``Mid-IR (any)'' in Table~\ref{tab:selres}). 
Interestingly, if we restrict the sample to mid-IR selected candidates that {\it were not}
also selected based on their variability or X-ray properties (``Mid-IR (only)'') the
overall yield is still 18\% (27\% bright).  This means that a large fraction of the mid-IR-selected
quasars are not being selected by the variability or X-ray criteria.  At least
for the latter, \cite{2009ApJ...696..891H} and \cite{2010ApJ...713..970A}
have previously noted that X-ray and mid-IR selection methods tend to select
different sources. 

\begin{figure*}
\centering
\includegraphics[width=15cm]{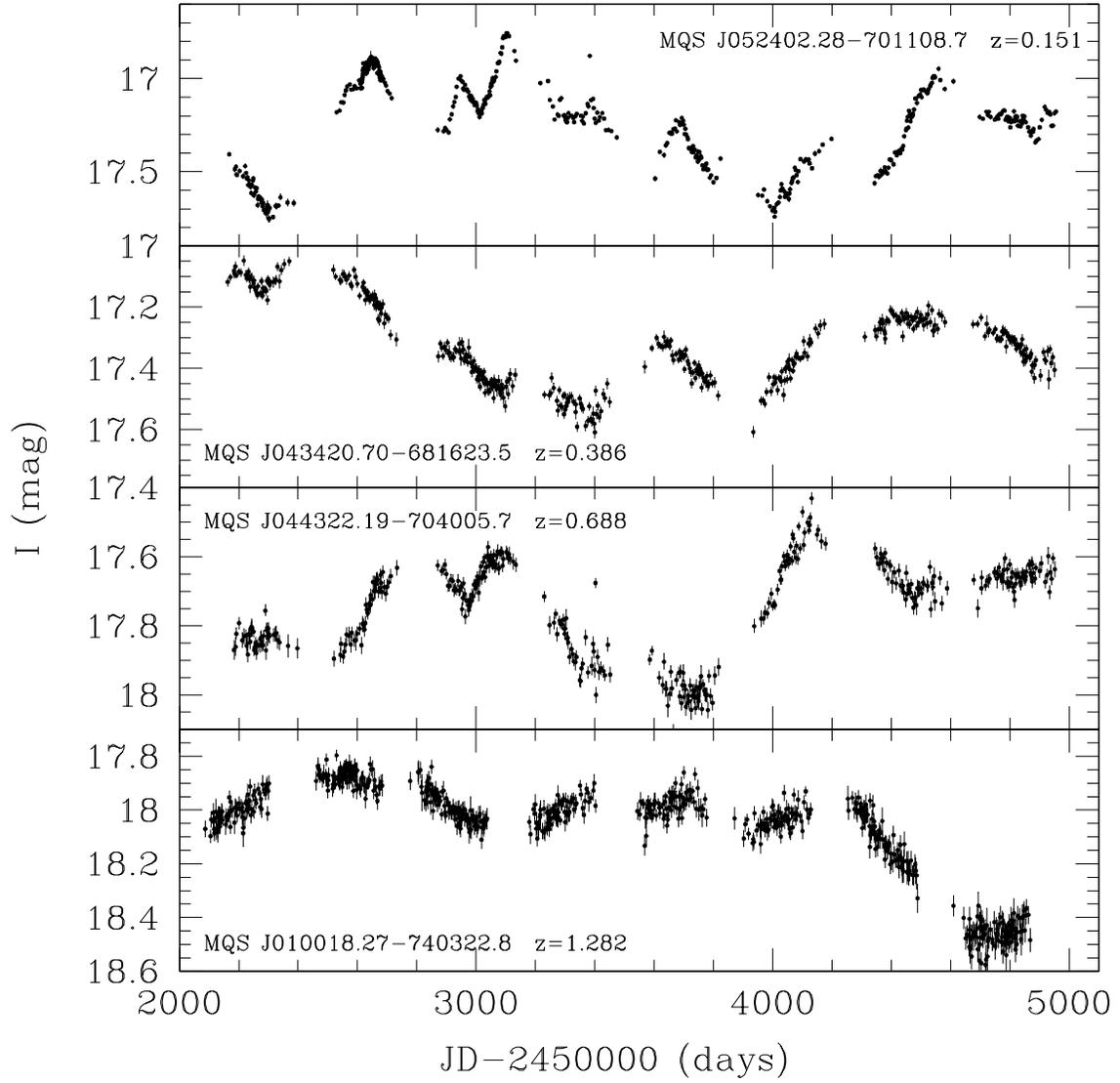}
\caption{Four examples of OGLE-III light curves for new MQS quasars (labeled in panels).} 
\label{fig:O3var}
\end{figure*}

We discuss the variability selection results in three parts.  First, we consider
variability selection as actually used to select candidates, and then we 
discuss the consequences of adding the tighter restrictions of either 
\cite{2010ApJ...708..927K} or \cite{2010ApJ...714.1194S}.
Figure~\ref{fig:O3var} shows four examples of the OGLE-III
light curves of newly identified quasars.

For variability selection as implemented, we started with 50 million MC 
sources.  After applying {\it Cuts~1} and {\it 2} ($\ln L_{\rm best}>\ln L_{\rm noise}+2$ and 
$16.0/16.5<I<19.5$~mag), 680 thousand possibly variable sources remained.  Adding
the restriction on DRW time scales ($1 < \log(\tau/{\rm days}) < 5$) reduced this
to 260 thousand sources, and then only 37 thousand sources remained after 
{\it Cuts~4} and {\it 5} ($0.1<\gamma<0.9$; $A<0.4$~mag). There are still large
numbers of false positives, primarily ``ghost variables'' where fainter stars pick
up a variability signal because they lie in the extended PSF wings of 
brighter variable stars.  After visually inspecting this final list we were left
with the $\sim$1400 real candidates.  The resulting efficiency is quite good,
with 34\% of these variability-selected candidates confirmed as AGN.  

\begin{figure*}
\centering
\includegraphics[width=12.5cm]{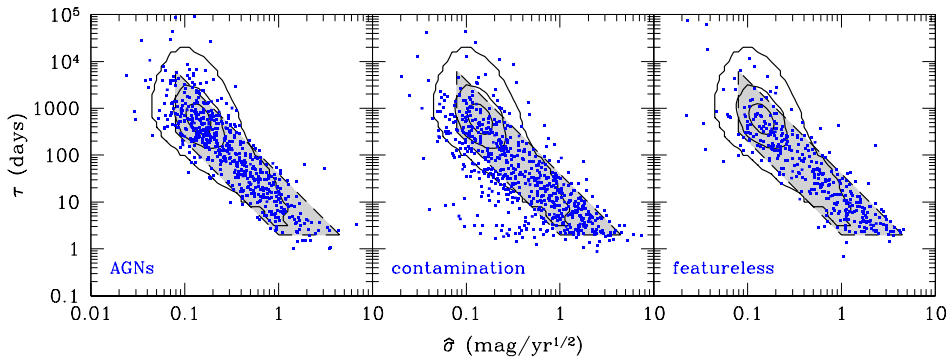}
\caption{$\tau$-$\hat{\sigma}$ (time scale-modified amplitude) variability plane as defined in \cite{2010ApJ...708..927K}.
In the left, middle, and right panel, we show MQS AGNs, contaminating sources, and objects with featureless spectra, respectively.
We show the \cite{2010ApJ...708..927K} trapezoid AGN selection region (gray area) 
and density contours (1, 10, and 20 per 0.1 dex bins in both axes) 
for $\sim$9000 variable SDSS AGNs from \cite{2010ApJ...721.1014M}.
The \cite{2010ApJ...708..927K} cut was designed to return high purity samples given the variability properties of contaminating stars. 
We extended this selection region (see Section~\ref{sec:AGNsel}) to probe the $\tau$-$\hat{\sigma}$ variability plane. 
The trapezoid contains 77\% of the variability-selected confirmed AGNs and 59\% of all confirmed AGNs.} 
\label{fig:tau_sigma}
\end{figure*}

We cannot retrospectively impose the exact selection procedures we introduced in
\cite{2010ApJ...708..927K} because of the additional selection cuts we introduced in Paper II 
 and continued to use here.  We can, however, examine the effects of
the additional restrictions on $\tau$ and $\sigma$ from \cite{2010ApJ...708..927K} 
on the present sample, as shown in Figure~\ref{fig:tau_sigma}.
It shows the distribution of our confirmed AGNs in the space for the DRW
parameters along with the selection region proposed in \cite{2010ApJ...708..927K}.
A very high fraction (77\%) of all variability-selected MQS quasars (59\% of all confirmed AGNs) lie in this 
narrower selection region, as we would also expect given the parameter distribution of 
the SDSS Stripe 82 quasars from \cite{2010ApJ...721.1014M}.  
If we apply the remaining cuts from \cite{2010ApJ...708..927K} on
the variability amplitude as a function of magnitude but not the cuts on the ratio
 of the $V$- and $I$-band variability amplitudes, 74\% of the sample remains (58\% of all confirmed AGNs).    
The level of contamination seen in 
Figure~\ref{fig:tau_sigma} looks higher than in \cite{2010ApJ...708..927K} because
there we only showed the distribution of other variable sources from the $\sim 2$~deg$^2$
analyzed for variability by OGLE-II (\citealt{1997AcA....47..319U}) rather than 
the OGLE-III sample (\citealt{2008AcA....58...69U}).
Overall, the yield for variable sources (Equation~\ref{eqn}) is $\sim 34\%$ (``Var (any)''), but in this restricted 
region of the $\tau$-$\hat{\sigma}$ plane (also using the remaining \citealt{2010ApJ...708..927K} 
cuts), it is $\sim 45\%$ (``Var (any)$+$DRW'') and by
definition these are all bright $16.0/16.5<I<19.5$~mag sources.
Almost all the confirmed variability-selected AGNs were also selected as
mid-IR candidates, probably because all they were also required to be
relatively bright.  As a result, the yield for those that were not also selected as
either X-ray or mid-IR candidates is low (8\% for ``Var. (only)'').  

\begin{figure*}
\centering
\includegraphics[width=15cm]{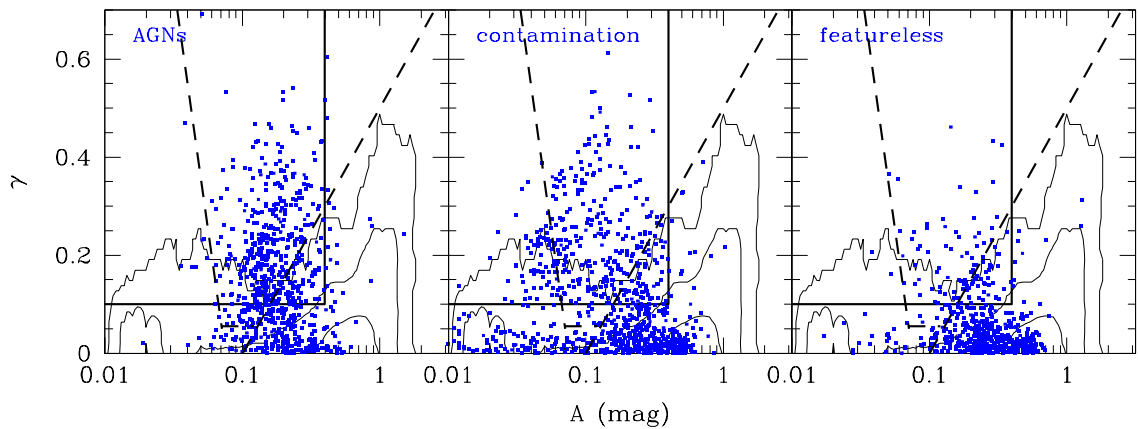}
\caption{$A$-$\gamma$ (amplitude-structure function slope) variability plane 
as in \cite{2010ApJ...714.1194S}.  In the left, middle, and right panel, we 
show MQS AGNs, contaminating, and objects with featureless spectra, respectively.
The dashed lines correspond to \cite{2010ApJ...714.1194S} AGN selection region 
(above these lines), and the solid vertical ($A=0.4$~mag) and solid horizontal ($\gamma=0.1$) lines
are our AGN selection cuts (above horizontal and to the left of vertical lines).
The contours are for a 1~deg$^2$ area with a typical LMC stellar density that contains 30000 objects.
The objects are counted in $\Delta A=0.02$ dex and $\Delta \gamma=0.02$ bins. 
The outer, middle, and inner contours are for 1, 10, and 100 objects per bin.  Objects
outside the selection regions were selected by other criteria (mid-IR and/or X-ray).
} 
\label{fig:schmidt}
\end{figure*}

Similarly, we can use the narrow variability selection criteria on $A$ and
$\gamma$ based on the structure function approach from \cite{2010ApJ...714.1194S}.
As shown in Figure~\ref{fig:schmidt}, they used the criteria that
$\gamma>0.5\times\log_{10}(\rm A)+0.50$, 
$\gamma>-2\times\log_{10}(\rm A)-2.25$, and $\gamma>0.055$.
For these tighter criteria, 48\% of the variability-selected sources were confirmed
to be AGNs and 41\% (83\%) of the confirmed (and variability-selected) AGNs satisfy the criteria. 
The 83\% is high because our variability selection method was quite
similar to the full procedures from \cite{2010ApJ...714.1194S}.
While they never contemplated using their 
method in dense stellar fields, it works reasonably well. 

\begin{figure}
\centering
\includegraphics[width=12.5cm]{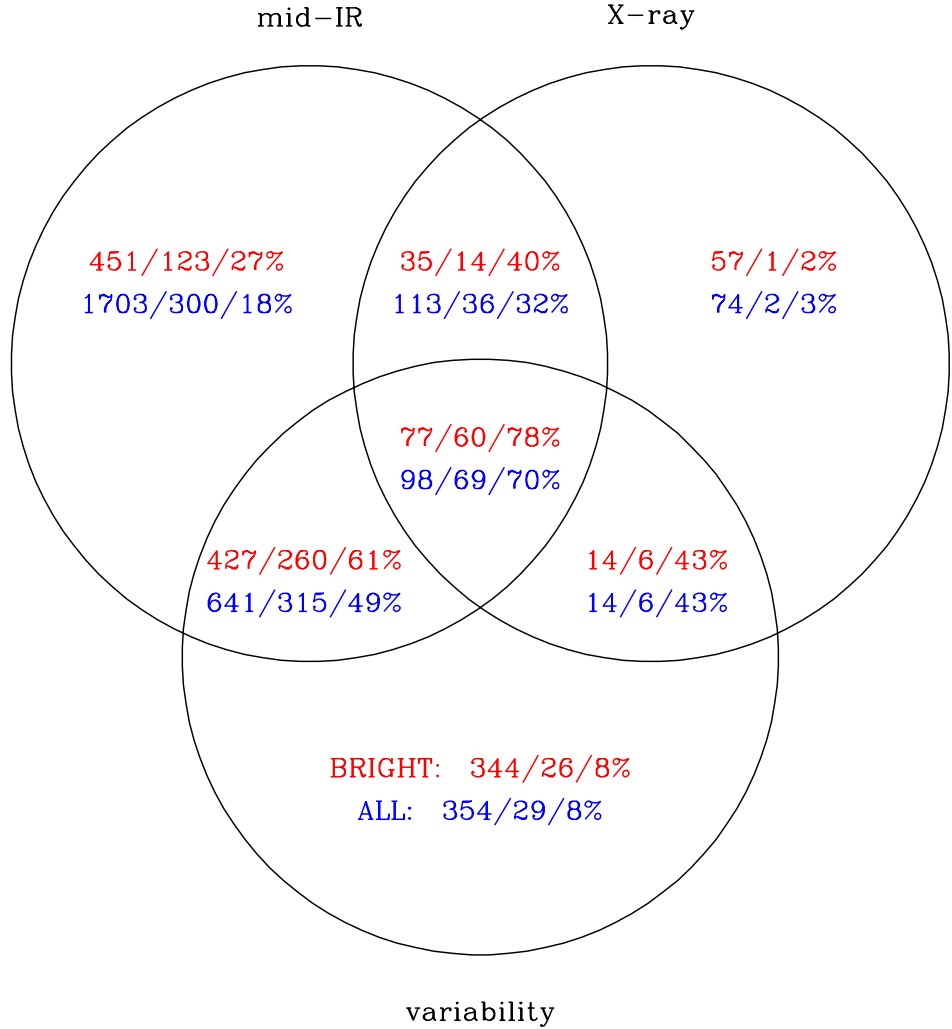}
\caption{
The Venn diagram for the confirmed AGNs, showing efficiencies of the three AGN selection methods. 
The numbers on the left (middle, right) are the numbers of observed targets (confirmed AGNs, yields).
The upper (red) numbers are for bright ($I<19.5$~mag) sources and the bottom ones (blue) are for all sources.  
See Table~\ref{tab:selres} for details.}
\label{fig:VennConf}
\end{figure}

\cite{2012ApJ...747..107K} selected 2566 AGN candidates spread over roughly
$40$~deg$^2$ behind the LMC based on their 
optical variability in the MACHO survey and then reduced the sample to 663 ``high quality''
candidates based on their optical, mid-IR, and/or X-ray properties.  Although the
MACHO sample is brighter, with a median magnitude of $\sim18.2$~mag rather than
our $19.6$~mag, the \cite{2012ApJ...747..107K} sample has a significantly
 higher surface density of 31~candidates/deg$^2$
as compared to 13~candidates/deg$^2$ for the MQS sample at the same magnitude
limit.  For comparison, the expected surface density of quasars brighter than
$18.2$~mag is only 3.2~quasars/deg$^2$, which means that the contamination levels
in the \cite{2012ApJ...747..107K} variability selected sample are significantly
higher than for the MQS samples, with upper limits on the purities of the
\cite{2012ApJ...747..107K} and MQS variability-selected samples of order
 10\% and 25\% respectively. The surface density of the ``high quality'' sample is
much lower, and in fact drops below the expected surface density of quasars at
fainter magnitudes, indicating that it must be substantially incomplete even
if it has little contamination.  There
are 248 (216) matches of their sample (high quality subset) to our MQS
samples for a matching radius of $3\farcs0$ with 133 (131) being confirmed
quasars.   \cite{2012ApJ...747..107K} attempt to compare their selection
methods to ours by contrasting the 131 MQS quasars in the sample of 248
candidates matched to their full sample (61\%) to the 7\% MQS yield (Paper~II) for
variability-selected quasars that were neither X-ray nor mid-IR-selected. 
Even if there was an independent spectroscopic follow-up of the \cite{2012ApJ...747..107K}
sample, one would need to either compare samples selected based only
on variability ($131/216=61\%$ versus 34\% for MQS) or variability-selected
samples not selected by other methods ($2/32=6\%$ versus 8\% for MQS)
rather than mixing the two possibilities.
More fundamentally, unless the \cite{2012ApJ...747..107K} selection methods
are completely devoid of any new information on whether sources are quasars, 
the apparent efficiency of the \cite{2012ApJ...747..107K} sub-sample 
contained in the MQS sample must be higher than the efficiency of the
MQS sample as a whole.  In essence, \cite{2012ApJ...747..107K} are
adding a fourth selection method and then comparing the completeness
of the intersection of (say) selection methods $2$+$4$ to the completeness of
selection method $2$ alone. This holds even if both methods are variability
selection methods, either independent statistics applied to the same data
set or (as in this case) different statistics applied to two different 
data sets.  As we see from Figure~\ref{fig:VennConf} and Table~\ref{tab:selres} the completeness obtained
from the intersections of selection methods are always markedly higher than
those for one method alone.   Without an independent spectroscopic
study of the \cite{2012ApJ...747..107K} sample it is impossible to compare
the efficiency of the different selection methods beyond the crude comparison
of the surface density of candidates to the surface density of quasars
discussed above.

Finally, Table~\ref{tab:selres} shows the effects of using various 
combinations of the selection methods. 
For example, samples that combine mid-IR$+$variability,
mid-IR$+$X-ray and variability$+$X-ray have yields of 52\% (63\%), 49\% (65\%), 
and 66\% (71\%) for all (bright) sources. In
these results we include objects independent of their status based on the
third selection method.  If objects are selected by two methods and not
by the third, the yields are generally significantly lower, at 
49\% (61\%), 32\% (40\%), and 43\% (43\%), respectively.  
The various possible overlapping selection
choices are graphically illustrated as a Venn diagram in Figure~\ref{fig:VennConf}.  

%%%%%%%%%%%%%%%%%%%%%  SUMMARY  %%%%%%%%%%%%%%%%%%%%

\section{Summary}
\label{sec:summary}

In this paper, we report the final spectroscopically confirmed AGN sample from {\it The Magellanic Quasars Survey} -- the largest 
spectroscopic search for MC quasars to date.  We obtained spectra for 2248 (766) LMC (SMC) sources 
and identified \allQSOsL\ (\allQSOsS) as AGNs. We also confirmed \knoQSOsL\ (\knoQSOsS) known LMC 
(SMC) AGNs and were unable to confirm one.  The total number of confirmed MQS quasars is
\allQSOs, of which \newQSOs\ are new.   Thus, the MQS has increased the number of quasars
known behind the MCs by an order of magnitude to an overall total of roughly 800 quasars.
This provides a dense network of proper motion reference points for improving measurements of
the internal and bulk proper motions of the MCs (e.g., \citealt{2013ApJ...764..161K,2013arXiv1305.4641V}), 
and these are the quasars with the best long-term, densely sampled light curves for studying 
quasar variability physics (e.g., \citealt{2009ApJ...698..895K,2010ApJ...721.1014M}).  
Also, 50 quasars brighter than $I \lesssim 18$~mag enable studies of the absorption by the 
ISM/IGM. We roughly estimate that we have achieved $\sim 75\%$ completeness for $I<19$~mag
quasars in the OGLE-III regions of the MCs.

The nature of the AAOmega instrument, with many more fibers than needed given the numbers of quasars
brighter than our effective magnitude limit of $I\approx20.5$~mag, means that we also obtain spectra
of many contaminating LMC sources.  Because we only target sources that have ``abnormal''
properties for stars, the contaminating sources are a mixture of dusty or accreting sources,
including many YSOs, PNe and Be stars.  These sources will be discussed elsewhere.  Despite
fully populating the fibers, the yields from the various selection methods are quite good,
particularly when combined. Individual methods typically have yields of order 30\%,
combinations of two methods have yields of order 55\% and combining all three has a yield
of 70\%.  Of course, the number of available targets also declines, and the overall 
number of AGN identified by only one, two or all three methods is 331, 357, and 69, respectively,
because of the usual trade-offs between completeness and contamination.  
In \cite{2009ApJ...701..508K} and \cite{2010ApJ...708..927K} we argued that mid-IR and
variability selection methods would be effective despite the high stellar densities of
the MCs, and the MQS provides excellent confirmation.  Since OGLE-III covered the densest regions
of the MCs, expanding the search for quasars to the larger OGLE-IV region will be
significantly easier because of the reduced stellar densities.  Doing so, however, requires
somewhat longer term OGLE-IV light curves to carry out the variability selection since
the mid-IR and X-ray surveys of the MCs do not extend over the much larger OGLE-IV
survey regions.

%%%%%%%%%%%%%%%%%%%%%  ACKNOWLEDGMENTS  %%%%%%%%%%%%%%%%%%%%

\acknowledgments

This research is based on observations made with the Anglo-Australian Telescope, 
for which the observing time was granted by the Optical Infrared Coordination Network for Astronomy (OPTICON).
This research has made use of the SIMBAD database, operated at CDS, Strasbourg, France. 
This research has also made use of the NASA/IPAC Extragalactic Database (NED) which is operated by the Jet Propulsion Laboratory (JPL), 
California Institute of Technology (Caltech), under contract with the National Aeronautics and Space Administration (NASA). 
C.S.K. is supported by NSF grant AST-1009756.
The OGLE is supported by the European Research Council under 
the European Community's Seventh Framework Programme (FP7/2007-2013),
ERC grant agreement no. 246678 to A.U. 
The work in this paper was partially supported by the Polish
Ministry of Science and Higher Education through the program ''Ideas
Plus'' award No. IdP2012 000162 to I.S.

%%%%%%%%%%%%%%%%%%%%%  BIBLIOGRAPHY  %%%%%%%%%%%%%%%%%%%%

%%%%%%%%%%%%%%%%%%%% FIGURE %%%%%%%%%%%%%%%%%%%%
\clearpage

%%%%%%%%%%%%%%%%%%%% FIGURE %%%%%%%%%%%%%%%%%%%%
\clearpage

%%%%%%%%%%%%%%%%%%%% FIGURE %%%%%%%%%%%%%%%%%%%%
\clearpage

%%%%%%%%%%%%%%%%%%%% FIGURE %%%%%%%%%%%%%%%%%%%%
\clearpage

%%%%%%%%%%%%%%%%%%%% FIGURE %%%%%%%%%%%%%%%%%%%%
\clearpage

%%%%%%%%%%%%%%%%%%%% FIGURE %%%%%%%%%%%%%%%%%%%%
\clearpage

%%%%%%%%%%%%%%%%%%%% FIGURE %%%%%%%%%%%%%%%%%%%%
\clearpage

%%%%%%%%%%%%%%%%%%%% FIGURE %%%%%%%%%%%%%%%%%%%%
\clearpage

%%%%%%%%%%%%%%%%%%%% FIGURE %%%%%%%%%%%%%%%%%%%%
\clearpage

%%%%%%%%%%%%%%%%%%%% FIGURE %%%%%%%%%%%%%%%%%%%%
\clearpage

%%%%%%%%%%%%%%%%%%%% TABLE %%%%%%%%%%%%%%%%%%%%
\clearpage

\begin{deluxetable}{lcccccr}
\tablecaption{The MQS Observing Log. \label{tab:MQSlog}}
\tablehead{Field & R.A. & Decl. & $N_{\rm cand}$ & $N_{\rm QSO}$ & $T_{\rm exp}$ (hr) & Paper}
\startdata
LMC1 & 04:41:43 & $-$69:50:29 & 217 & 45 & 1.5 & III \\
LMC2 & 04:43:19 & $-$68:19:18 & 193 & 71 & 1.5 & III \\
LMC3 & 04:56:52 & $-$67:07:48 & 209 & 53 & 1.5 & III \\
LMC4a & 05:00:51 & $-$70:27:49 & 221 & 41 & 0.4 & II \\
LMC4b & 05:00:51 & $-$70:27:49 & 216 & 59 & 1.5 & III \\
LMC5 & 05:01:42 & $-$68:49:26 & 265 & 75 & 1.5 & III \\
LMC6 & 05:14:27 & $-$67:35:00 & 189 & 46 & 1.9 & III \\
LMC7 & 05:19:43 & $-$69:31:07 & 307 & 36 & 1.5 & II\\
LMC8 & 05:21:54 & $-$71:02:48 & 247 & 60 & 1.0 & II\\
LMC9 & 05:32:54 & $-$68:31:01 & 220 & 61 & 1.5 & III \\
LMC10 & 05:41:31 & $-$71:36:00 & 210 & 37 & 0.7 & II\\
LMC11 & 05:41:56 & $-$70:11:07 & 263 & 55 & 1.5 & III \\
LMC12 & 05:52:39 & $-$68:57:51 & 180 & 56 & 1.5 & III \\
\hline
SMC PILOT & 00:52:00 & $-$72:48:00 & 268 & 32 & 1.5 & I \\
SMC1 & 00:33:45 & $-$73:25:07 & $\cdots$ & $\cdots$ & $\cdots$ & $\cdots$ \\
SMC2 & 00:49:10 & $-$73:51:48 & 273 & 53 & 1.5 & III \\
SMC3 & 00:55:09 & $-$72:15:13 & $\cdots$ & $\cdots$ & $\cdots$ & $\cdots$ \\
SMC4 & 01:14:18 & $-$72:00:43 & 256 & 67 & 1.5 & III \\
SMC5 & 01:15:12 & $-$73:33:57 & 239 & 76 & 1.5 & III
\enddata
\tablecomments{Each field has a 1 deg radius. $N_{\rm QSO}$ does not have to add up to \newQSOs\ 
new QSOs and \knoQSOs\ known QSOs because the fields overlap slightly (see 
Figure~\ref{fig:MQS_fields}) and a quasar can be observed in several fields.}
\end{deluxetable}

%%%%%%%%%%%%%%%%%%%% TABLE %%%%%%%%%%%%%%%%%%%%
\clearpage

\begin{landscape}
\begin{deluxetable}{l|cc|cc|cc|cc|rr|cc}
\tabletypesize{\scriptsize}
\tablecaption{MQS Quasars Behind the LMC\label{tab:results_LMC}. }
\tablewidth{0pt}
\tablehead{MQS AGN Name & R.A. & Decl. & z & $\mu$ & $V$ & $I$ & $A_V$ & $A_I$ & $K_V$ & $K_I$ & $M_V$ & $M_I$  \\
 & (deg) & (deg) & & (mag) & (mag) & (mag) &  (mag) & (mag) & (mag) & (mag) & (mag) & (mag) \\
(1) & (2) & (3) & (4) & (5) & (6) & (7) & (8) & (9) & (10) & (11) & (12) & (13) }
\startdata
MQS J043110.08$-$695241.5 & 67.792000 & $-$69.878194 & 1.548 & 45.19 & 19.63 & 18.89 & 0.29 & 0.17 & $-$0.40 & $-$0.07 & $-$25.45 & $-$26.40 \\
MQS J043151.34$-$692437.9 & 67.963917 & $-$69.410528 & 0.594 & 42.60 & 19.77 & 19.01 & 0.22 & 0.13 & $-$0.21 &    0.18 & $-$22.84 & $-$23.90 \\
MQS J043200.60$-$693846.5 & 68.002500 & $-$69.646250 & 1.409 & 44.93 & 20.74 & 19.89 & 0.22 & 0.13 & $-$0.38 & $-$0.01 & $-$24.03 & $-$25.17 \\ 
MQS J043221.19$-$701129.5 & 68.088292 & $-$70.191528 & 0.957 & 43.90 & 17.62 & 17.26 & 0.29 & 0.17 & $-$0.42 &    0.23 & $-$26.15 & $-$27.04 \\
MQS J043232.77$-$694433.2 & 68.136542 & $-$69.742556 & 2.162 & 46.07 & 19.01 & 18.15 & 0.22 & 0.13 & $-$0.59 & $-$0.19 & $-$26.69 & $-$27.87 \\
MQS J043238.16$-$700438.4 & 68.159000 & $-$70.077333 & 2.145 & 46.06 & 19.73 & 18.91 & 0.22 & 0.13 & $-$0.59 & $-$0.19 & $-$25.96 & $-$27.09 \\
MQS J043259.64$-$693653.0 & 68.248500 & $-$69.614722 & 0.948 & 43.87 & 19.79 & 19.28 & 0.22 & 0.13 & $-$0.42 &    0.23 & $-$23.89 & $-$24.95 \\ 
MQS J043308.66$-$701341.5 & 68.286083 & $-$70.228194 & 1.428 & 44.97 & 17.91 & 17.13 & 0.29 & 0.17 & $-$0.38 & $-$0.01 & $-$26.97 & $-$28.00 \\ 
MQS J043322.97$-$680832.9 & 68.345708 & $-$68.142472 & 0.937 & 43.84 & 20.89 & 20.66 & 0.12 & 0.07 & $-$0.41 &    0.23 & $-$22.66 & $-$23.49 \\ 
MQS J043330.96$-$690844.0 & 68.379000 & $-$69.145556 & 3.028 & 46.96 & 19.28 & 18.72 & 0.24 & 0.14 & $-$0.77 & $-$0.27 & $-$27.16 & $-$28.11
\enddata
\tablecomments{The error code for magnitudes, reflecting no measurement, is 99.99. 
In column (5), we show the distance modulus $\mu=5\log(D_L/\rm{Mpc})+25$,
in (8) and (9) extinctions, in (10) and (11) $K$-corrections, and in (12) and (13) absolute magnitudes.
In column (19), N stands for a new AGN, II means an AGN reported in Paper~II, and K is for already known AGN.
(This table is available in its entirety in a machine-readable form in
the online journal. A portion is shown here for guidance regarding its
form and content.)}
\end{deluxetable}

\addtocounter {table} {-1}

\begin{deluxetable}{l|c|c|ccc|c|c|r}
\tabletypesize{\scriptsize}
\tablecaption{Continuation.}
\tablewidth{0pt}
\tablehead{MQS AGN Name &  OGLE-III & KK09 & Mid-IR & X-ray & Var. & Notes & Quality & Emission Lines \\
 & ID & Class & & & & & Flag & \\
& (14) & (15) & (16) & (17) & (18) & (19) & (20) & (21)}
\startdata
MQS J043110.08$-$695241.5 &  lmc157.7.2453 & QSO-Aa & 1 & 0 & 1 & N & Q2 & CIII], MgII      \\  
MQS J043151.34$-$692437.9 &  lmc156.8.4036 & QSO-Aa & 1 & 0 & 1 & N & Q1 & MgII, [OII], H$\beta$, [OIII]	 \\
MQS J043200.60$-$693846.5 &  lmc157.5.3272 & QSO-Aa & 1 & 0 & 1 & N & Q2 & MgII	\\
MQS J043221.19$-$701129.5 &   lmc158.4.202 & YSO-Aa & 1 & 0 & 1 & N & Q3 & MgII	\\
MQS J043232.77$-$694433.2 &   lmc157.3.244 & QSO-Aa & 1 & 0 & 1 & N & Q1 & Ly$\alpha$, SiIV, CIV, CIII]	\\
MQS J043238.16$-$700438.4 &   lmc157.1.247 & QSO-Aa & 1 & 0 & 1 & N & Q1 & Ly$\alpha$, SiIV, CIV, CIII]	\\
MQS J043259.64$-$693653.0 &   lmc157.4.758 & QSO-Aa & 1 & 0 & 1 & N & Q2 & MgII, [OII]      \\
MQS J043308.66$-$701341.5 &  lmc158.4.2814 & YSO-Aa & 1 & 0 & 1 & N & Q1 & CIII], MgII      \\
MQS J043322.97$-$680832.9 &  lmc154.7.1692 & QSO-Aa & 1 & 0 & 0 & N & Q3 & MgII	\\
MQS J043330.96$-$690844.0 &  lmc156.6.4143 & QSO-Aa & 1 & 0 & 1 & N & Q3 & Ly$\alpha$, SiIV, CIV
\enddata
\end{deluxetable}

%%%%%%%%%%%%%%%%%%%% TABLE %%%%%%%%%%%%%%%%%%%%

\begin{deluxetable}{l|cc|cc|cc|cc|rr|cc}
\tabletypesize{\scriptsize}
\tablecaption{MQS Quasars Behind the SMC\label{tab:results_SMC}. }
\tablewidth{0pt}
\tablehead{MQS AGN Name & R.A. & Decl. & z & $\mu$ & $V$ & $I$ & $A_V$ & $A_I$ & $K_V$ & $K_I$ & $M_V$ & $M_I$  \\
 & (deg) & (deg) & & (mag) & (mag) & (mag) &  (mag) & (mag) & (mag) & (mag) & (mag) & (mag) \\
(1) & (2) & (3) & (4) & (5) & (6) & (7) & (8) & (9) & (10) & (11) & (12) & (13) }
\startdata
MQS J003704.67$-$732229.6 &   9.269458 & $-$73.374889 & 0.750 & 43.24 & 18.89 & 18.49 & 0.05 & 0.03 & $-$0.32 &    0.18 & $-$24.08 & $-$24.96 \\
MQS J003857.54$-$741000.9 &   9.739750 & $-$74.166917 & 2.692 & 46.65 & 18.41 & 17.75 & 0.12 & 0.07 & $-$0.69 & $-$0.16 & $-$27.67 & $-$28.81 \\
MQS J003942.32$-$732428.1 &   9.926333 & $-$73.407806 & 0.382 & 41.46 & 20.08 & 19.10 & 0.07 & 0.04 & $-$0.02 &    0.23 & $-$21.43 & $-$22.63 \\
MQS J003947.82$-$743444.8 &   9.949250 & $-$74.579111 & 1.810 & 45.60 & 18.43 & 17.55 & 0.12 & 0.07 & $-$0.52 & $-$0.17 & $-$26.77 & $-$27.95 \\
MQS J003957.65$-$730603.6 &   9.990208 & $-$73.101000 & 0.569 & 42.51 & 19.85 & 19.43 & 0.10 & 0.06 & $-$0.20 &    0.18 & $-$22.56 & $-$23.33 \\
MQS J004023.71$-$741013.9 &  10.098792 & $-$74.170528 & 0.623 & 42.74 & 19.29 & 18.66 & 0.12 & 0.07 & $-$0.23 &    0.17 & $-$23.34 & $-$24.32 \\
MQS J004143.75$-$731017.1 &  10.432292 & $-$73.171417 & 0.217 & 40.10 & 21.55 & 20.96 & 0.12 & 0.07 &    0.01 & $-$0.03 & $-$18.68 & $-$19.18 \\
MQS J004145.04$-$725435.9 &  10.437667 & $-$72.909972 & 0.267 & 40.60 & 20.19 & 19.05 & 0.12 & 0.07 &    0.01 & $-$0.01 & $-$20.55 & $-$21.62 \\ 
MQS J004152.35$-$735626.8 &  10.468125 & $-$73.940778 & 0.422 & 41.72 & 21.57 & 20.30 & 0.10 & 0.06 & $-$0.06 &    0.24 & $-$20.19 & $-$21.72 \\
MQS J004241.66$-$734041.3 &  10.673583 & $-$73.678139 & 0.905 & 43.76 & 20.18 & 19.91 & 0.10 & 0.06 & $-$0.40 &    0.23 & $-$23.28 & $-$24.14
\enddata
\tablecomments{The error code for magnitudes, reflecting no measurement, is 99.99. In column (5), 
we show the distance modulus $\mu=5\log(D_L/\rm{Mpc})+25$, in (8) and (9) extinctions, in (10) and (11) $K$-corrections, 
and in (12) and (13) absolute magnitudes. In column (19), N stands for a new AGN, 
I means an AGN reported in Paper~I, and K is for already known AGN.
(This table is available in its entirety in a machine-readable form in
the online journal. A portion is shown here for guidance regarding its
form and content.)}
\end{deluxetable}

\addtocounter {table} {-1}

\begin{deluxetable}{l|c|c|ccc|c|c|r}
\tabletypesize{\scriptsize}
\tablecaption{Continuation.}
\tablewidth{0pt}
\tablehead{MQS AGN Name &  OGLE-III & KK09 & Mid-IR & X-ray & Var. & Notes & Quality & Emission Lines \\
 & ID & Class & & & & & Flag & \\
& (14) & (15) & (16) & (17) & (18) & (19) & (20) & (21)}
\startdata
MQS J003704.67$-$732229.6 & smc130.2.11076 & QSO-Aa & 1 & 0 & 1 & N & Q2 & MgII    \\
MQS J003857.54$-$741000.9 &   smc128.8.594 & QSO-Aa & 1 & 0 & 1 & N & Q1 & Ly$\alpha$, SiIV, CIV, CIII]     \\
MQS J003942.32$-$732428.1 &  smc125.7.5747 & QSO-Aa & 1 & 0 & 1 & N & Q1 & [OII], H$\beta$, [OIII]	       \\
MQS J003947.82$-$743444.8 &  smc129.7.2762 & QSO-Aa & 1 & 0 & 0 & N & Q1 & CIV, CIII], MgII      \\
MQS J003957.65$-$730603.6 &  smc125.5.6063 & QSO-Aa & 1 & 1 & 1 & I & Q2 & MgII, [OIII]   \\
MQS J004023.71$-$741013.9 &  smc128.8.9401 & QSO-Aa & 1 & 0 & 0 & N & Q1 & MgII    \\
MQS J004143.75$-$731017.1 & smc125.5.18504 & QSO-Aa & 1 & 0 & 0 & N & Q2 & [OII], [OIII]   \\
MQS J004145.04$-$725435.9 & smc126.8.16111 & QSO-Aa & 1 & 1 & 0 & I & Q1 & [OII], H$\beta$, [OIII]	      \\ 
MQS J004152.35$-$735626.8 &  smc128.2.2551 & QSO-Aa & 1 & 0 & 0 & N & Q3 & [OII], [OIII]   \\
MQS J004241.66$-$734041.3 &  smc128.4.5190 & QSO-Aa & 1 & 1 & 0 & N & Q2 & MgII  
\enddata
\end{deluxetable}

\end{landscape}

%%%%%%%%%%%%%%%%%%%% TABLE %%%%%%%%%%%%%%%%%%%%
\clearpage

\begin{deluxetable}{l|ccc|ccc}
\tablewidth{\columnwidth}
\tabletypesize{\small}
\tablecaption{MQS yields\label{tab:selres}.}
\tablehead{Selection  & \multicolumn{3}{c}{All MC sources} & \multicolumn{3}{c}{$I<19.5$ mag MC sources} \\
\hline
 & Observed & Confirmed & Weighted & Observed & Confirmed & Weighted \\
 & targets & AGNs & Yield \% & targets & AGNs & Yield \%}
\startdata
Mid-IR QSO-Aa & 2127 & 636  & 29           & 806 & 401 & 49 \\
Mid-IR QSO-Ab & 36   & 4    & 11           & 9   & 0   & 0  \\
Mid-IR QSO-Ba & 219  & 55   & 24 	   & 73  & 38  & 51 \\
Mid-IR QSO-Bb & 2    & 0    & 0 	   & 0   & 0   & $\cdots$ \\
Mid-IR YSO-Aa & 99   & 17   & 15 	   & 81  & 14  & 14 \\
Mid-IR YSO-Ab & 40   & 3    & 7 	   & 5   & 0   & 0  \\
Mid-IR YSO-Ba & 16   & 4    & 27 	   & 13  & 3   & 25 \\
Mid-IR YSO-Bb & 15   & 0    & 0 	   & 2   & 0   & 0  \\
\hline
Mid-IR (any)         & 2555 & 721 & 27  & 990 & 457 & 44 \\
X-ray (any)          & 299  & 113 & 30 	& 183 & 81  & 33 \\
Var. (any)           & 1107 & 419 & 34 	& 862 & 352 & 36 \\
Var (any)$+$DRW      & 513  & 226 & 45  & 513 & 226 & 45 \\
X-ray + Mid-IR (any) & 211  & 105 & 49  & 112 & 74  & 65 \\
Mid-IR + Var. (any)  & 739  & 384 & 52  & 504 & 320 & 63 \\
Var. + X-ray (any)   & 112  &  75 & 66  &  91 &  66 & 71 \\
\hline
\multicolumn{7}{c}{Priority 7}\\
\hline
Mid-IR (only) & 1703 & 300 & 18 & 451 & 123 & 27 \\
X-ray (only)  & 74   & 2   & 3 	& 57  & 1   & 2  \\
Var. (only)   & 354  & 29  & 8  & 344 & 26  & 8  \\
\hline
\multicolumn{7}{c}{Priority 8}\\
\hline
X-ray + Mid-IR (only) & 113 & 36  & 32 & 35  & 14  & 40 \\
Mid-IR + Var. (only)  & 641 & 315 & 49 & 427 & 260 & 61 \\
Var. + X-ray (only)   & 14  & 6   & 43 & 14  & 6   & 43 \\
\hline
\multicolumn{7}{c}{Priority 9}\\
\hline
all three & 98 & 69 & 70 & 77 & 60 & 78
\enddata
\end{deluxetable}

%%%%%%%%%%%%%%%%%%%% TABLE %%%%%%%%%%%%%%%%%%%%

\end{document}